# WHERE THE TREES FALL: MACROECONOMIC FORECASTS FOR FOREST-RELIANT STATES

Andrew Crawley^, Adam Daigneault*, Jonathan Gendron+

## Abstract

Several key states in various regions of the U.S. have experienced recent sawtimber as well as pulp and paper mill closures, which raises an important policy question: how have and will key macroeconomic and industry specific indicators within the U.S. forest sector likely to change over time? This study provides empirical evidence to support forest-sector policy design by using a vector error correction (VEC) model to forecast economic trends in three major industries – forestry and logging, wood manufacturing, and paper manufacturing – across six of the most forest-dependent states found by the location quotient (LQ) measure: Alabama, Arkansas, Maine, Mississippi, Oregon, and Wisconsin. Overall, the results suggest a general decline in employment and the number of firms in the forestry and logging industry as well as the paper manufacturing industry, while wood manufacturing is projected to see modest employment gains. These results also offer key insights for regional policymakers, industry leaders, and local economic development officials: communities dependent on timber-based manufacturing may be more resilient than other forestry-based industries in the face of economic disruptions. Our findings can help prioritize targeted policy interventions and inform regional economic resilience strategies. We show distinct differences across forest-dependent industries and/or state sectors and geographies, highlighting that policies may have to be specific to each sector and/or geographical area. Finally, our VEC modeling framework is adaptable to other resource-dependent industries that serve as regional economic pillars such as mining, agriculture, and energy production offering a transferable tool for policy analysis in regions with similar economic structures.

**Code/Data Availability**: Github link **JEL Codes:** C53, L73, Q23

**Funding Statement**: This work is supported by the McIntire-Stennis, project award no. [MEO-41825], from the U.S. Department of Agriculture's National Institute of Food and Agriculture.

+ = Corresponding Author: Virginia Tech Department of Economics, 880 West Campus Drive, Blacksburg, VA 24061, USA, email: jegendron@vt.edu

^ = Coauthor: University of Maine School of Economics, 5782 Winslow Hall, Orono, ME, 04469, USA, email: andrew.crawley@maine.edu

* = Coauthor: University of Maine School of Forest Resources, 5755 Nutting Hall, Orono, ME, 04469, USA, email: adam.daigneault@maine.edu

## 1. Introduction

Forest-dependent industries have long served a crucial role in the economic vitality of many rural U.S. communities across various regions, often serving as foundational pillars of local employment and public revenue (Balogh, 2018). However, these industries have undergone significant structural changes, such as market consolidation, global supply chain shifts, technological shifts, and declining demand for traditional wood products such as newsprint and paper (Toppinen & Kuuluvainen, 2010, Woodall et al, 2011). The consequences ripple through these communities leading to mill closures, job losses, declining tax bases, and reduced economic resilience, particularly in states heavily reliant on the forest sector (Sherman & Daigneault, 2023). These disruptions pose an urgent question for policymakers and economic development officials: *how are key macroeconomic and sector-specific indicators likely to evolve in forest-reliant communities, and what interventions can improve resilience?*

This paper aims to support evidence-based forest sector policy by offering industry-specific, state-level economic forecasts using a cointegration-based econometric framework with six historically forestry-reliant states: Alabama, Arkansas, Maine, Mississippi, Oregon, and Wisconsin. Specifically, we apply a vector error correction (VEC) model (e.g., Parajuli et al., 2016; Hetemaki et al., 2004) to forecast employment, output, wages, and firm dynamics across three core forest sector industries: (1) forestry and logging (NAICS 113), which encompasses tree cultivation, harvesting, and timber transport; (2) wood manufacturing (NAICS 321); and (3) paper manufacturing (NAICS 322), both of which involve processing harvested timber in a mill into lumber and paper products, respectively (USDA, 2011). While the literature often uses terms such as "timber", "logging", and "forestry" interchangeably, this study distinguishes between these core subsectors to more precisely identify which components of the forest economy are most vulnerable or resilient to future economic shifts. To identify the most forest-dependent states, we first calculate location quotients (LQs) for all U.S. states with measurable forest industry activity. These LQs quantify each state's relative economic dependence on the forest sector and guide the selection of states most in need of targeted policy attention. This approach provides a generalizable foundation for identifying other natural resource-based dependent industries. We then apply state and industry VECs for each highly forest-dependent state in our study.

Generating forward-looking region-specific economic forecasts, our analysis helps policymakers, industry leaders, and local development officials by providing information and strategies to best anticipate and mitigate disruptions with actions such as labor planning, economic diversification strategies, and targeted support programs for forest- and other natural resource-dependent communities (Anderson et al., 2002). Importantly, our analysis highlights that the impacts of structural change are not uniform: some industries (i.e. wood manufacturing) show modest resilience, while others (i.e. forestry and paper manufacturing) are expected to contract, requiring different policy responses. Furthermore, this modeling framework is transferrable and can be used to evaluate similar risks in other resource-dependent sectors such as mining, agriculture, or energy (e.g., Michieka & Gearhart 2015; Jiang & Liu, 2014; Bekhet & Yusop, 2009). The same forecasting techniques applied here can be used to assess economic resilience in other sectors or regions facing resource dependency (such as mining, agriculture, or energy). These sectors often share similar vulnerabilities to technological change, price shocks, and regulatory shifts. Understanding where economic vulnerabilities are likely to emerge enables more strategic allocation of public resources and more effective long-term planning. These vulnerabilities have been long studied in empirical economic analyses, particularly in the contexts of forecasting, labor dynamics, policy evaluation, price analysis, and trade. Forecasting models are central to this work: they help decision-makers understand not just where industries are today, but where they are heading. In practice, regional forecasts often inform state-level budgeting, labor force planning, infrastructure investment, and policy prioritization for rural and resource-dependent areas (Pattanayak et al., 2002; Federal Reserve System, 2016). Although all forecasting methods involve a degree of measurement error and respective uncertainty, selecting the most appropriate econometric model that can handle the particular characteristics of the data (such as long-run cointegration between variables) enhances accuracy and validity. This enhancement accordingly greatly improves the usefulness of forecasts in real-world policy design and decision making.

Existing literature already documents the high economic sensitivity of forest-reliant communities; for example, in Maine mill openings and closures have immediate and measurable effects on local economies (Gendron et al., 2019). Oregon provides a similar example, with the forestry sector contributing up to $375 million in annual state and local tax revenues as early as 2005 (Associated Oregon Loggers, 2021). In Mississippi, forest-based industries generated $925.73 million in tax revenue in 2018 and have yielded over $1 billion annually in timber harvests since 1993 (Henderson & Munn, 2008; Tanger & Measells, 2020). These industries also support thousands of direct and indirect jobs throughout the state (Munn, 1997). Wood and paper manufacturing, while downstream from forestry, are similarly vulnerable to supply shocks, price fluctuations and mill openings/closures, given their reliance on harvested raw materials. These industries are particularly significant in Alabama, Arkansas, Maine, Mississippi, Wisconsin, and Oregon. These states stand to benefit most from evidence-based policymaking tools that anticipate and mitigate the economic consequences of sectoral disruptions and also provide targeted support to its affected communities. Alabama alone contributes an estimated $7.9 billion to the economy in value added output from the forest sector (Fickle, 2017). Arkansas ranks first among southern states and fourth nationally in lumber production, with a total economic impact exceeding $12 billion as early as 2001 (Pelkki, 2005). Oregon's lumber industry accounts for more than half of all manufacturing employment in the state and has led the nation in wood product production since 1938 (Cloughesy, 2021; Robbins, 2021). While Oregon is less reliant on paper manufacturing, this industry is a significant contributor to economic activity in Alabama, Arkansas, Maine, Mississippi, and Wisconsin –states that consistently rank among the top in U.S. paper production (Ballweg, 2017).

The remainder of this paper is organized as follows. Section 2 reviews the econometric literature on forest sector forecasting and related policy analysis. Section 3 describes the data and methodology, with an emphasis on the Vector Error Correction (VEC) model's capacity to analyze long-run relationships and forecast structural trends. Section 4 presents and interprets the results from a policy and resilience standpoint. Section 5 concludes with a discussion of actionable insights for forest sector stakeholders and opportunities to apply this approach to other resource-dependent economies that rely on industries such as mining, agriculture, and energy production.

## 2. Literature Review

Forecasting macroeconomic trends is a well-established approach for understanding and responding to real-world economic challenges. These forecasts are often used to support policy decisions aimed at stabilizing or stimulating key economic indicators such as employment and output (Anderson et al., 2002; Eicher et al., 2019; Poudel & Dahal, 2025). For example, the U.S. Federal Reserve routinely produces macroeconomic forecasts to guide monetary policy and assess broader economic conditions across the globe (Federal Reserve System, 2016). At more localized levels, similar forecasting methodologies have been applied to state or regional economies and to specific industries - including the forest sector – where shifts in output, employment, and firm activity can significantly affect the wellbeing of resource-dependent communities. In such contexts, policy intervention informed by accurate forecasts can help mitigate economic downturns and support community resilience.

Selecting an appropriate econometric model is crucial to generating reliable forecasts. An essential first step is determining whether the underlying data is stationary or non-stationary. Non-stationary (or cointegrated) data indicates that variables move with each other due to an underlying long-run relationship, whereas stationary data suggests that variables fluctuate independently over time (Engle & Granger, 1987). Mistaking these properties can result in model specification and accordingly unreliable inferences, undermining the reliability of policy recommendations (Hill et al., 2008). Even when short-run dynamics suggest independence between variables, long-run equilibrium relationships may exist and must be modeled accordingly as cointegrated. Since traditional stationary models cannot properly account for cointegrated relationships, more advanced econometric techniques have been developed including the autoregressive integrated moving average (ARIMA), error correction (EC) and vector error correction (VEC) models (Engle & Granger, 1987). Beyond these standard approaches, hybrid and unique models, such as behavioral adaptive expectation models, have also been explored (Haji-Othman, 1991). Formal statistical tests such as the Johansen test and the Augmented Dickey-Fuller (ADF) test for stationarity are widely used to guide model selection (Johansen, 1988; Al-Ballaa, 2005).

Table 1 summarizes common econometric methods used in the forest sector forecasting. EC models are often favored for their ability to handle cointegrated data. For instance, Polemis (2007) analyzed energy demand in Greece with an EC, and Song et al. (2011) used a stationary two-stage least squares (2SLS) model. In the forest sector, EC models have been utilized to analyze sawlog stumpage prices in the southern U.S. (Mei et al., 2010). However, EC models can only accommodate a single cointegrated equation, making them less flexible when modeling more complex systems like multiple interdependent variables (Hill et al., 2008; Lambert et al., 2013). In such cases, vector autoregression (VAR) and VEC models are capable of handling multiple cointegrated equations. Comparative studies have shown that VAR and VEC models outperform ARIMA and other specifications in terms of forecasting accuracy, particularly in contexts involving complex interrelationships between multiple economic indicators (Hetemaki et al., 2004; Malaty et al., 2006; Mei et al., 2010). In the forest sector, where employment, output, firm activity and wages often evolve together, VEC models offer a powerful framework for capturing both short-run policy-relevant fluctuations and long-run structural dynamics (Parajuli & Chang, 2015; Kolo & Tzanova, 2017; Hernandez-Diaz et al., 2024).

**Table 1.** Summary of forest sector econometric forecasting literature.

| Study | Region | Sector | Method | Timesteps | Results |
|---|---|---|---|---|---|
| Baek (2012) | U.S. & Canada | Softwood lumber imports (U.S. from Canada) | EC | 186, monthly | Identifies a long-run cointegrated relationship between U.S. lumber imports and selected macroeconomic variables. |
| Banaś & Banaś (2021) | Poland | Roundwood | Seasonal ARIMA | 64, quarterly | Results find that seasonality and addition of an exogenous variable substantially improve forecast accuracy. |
| Bhatt et al. (2021) | Nepal | Timber & non-timber products | ARIMA | 15, annually | Forecasts the economic contribution of the forestry sector finding highest revenue from timber and fuelwood. |
| Ferguene et al. (2023) | Quebec | Lumber demand | OLS, LASSO & RIDGE | 27, annually | The results point to OLS performing better than both LASSO and RIDGE model selection. |
| Gendron et al. (2019) | Maine (U.S.) | Forestry & Logging | VECM | 69, quarterly | Forecasts suggest stable output and prices, but declining local employment and firm counts in Maine's forestry sector. |
| Hetemaki et al. (2004) | Germany & Finland | Lumber exports & sawlog demand | VAR, ARIMA, VEC | 64, quarterly | Demonstrates improved forecasting accuracy using multivariate models (VAR and VEC) over ARIMA. |
| Hernandez-Diaz et al. (2024) | Southern U.S. | Sawtimber & pulpwood stumpage | Wavelet Analysis | 176, quarterly | Reveals a novel relationship between hurricane activity and stumpage price fluctuations in the Southern U.S. |
| Hietala et al. (2013) | Finland, Sweden & UK | Sawn softwood exports | VAR | 56, quarterly | Shows that exchange rate fluctuations (euros & pounds) significantly influence Finnish sawn softwood exports. |
| Kolo & Tzanova (2017) | Germany | Raw timber trade (imports & exports) | VECM | 68, quarterly | Estimates volumes and prices of Germany's raw timber exports and imports using a VECM framework. |
| Kozuch et al. (2023) | Poland | Roundwood | Artificial Neural Network | 64, quarterly | Artificial neural networks exhibit generally better fit, but results highlight that unique analysis is needed per species of wood. |
| Lamichhane et al. (2025) | Southern U.S. | Softwood lumber & timber prices | AR with Markhov Switching | 96, quarterly | Price divergences likely driven by external shocks and their changes, especially the housing market. |
| Malaty et al. (2006) | Finland | Sawlog markets | ARIMA & VAR | 120, monthly | Finds regional price asymmetries in Finnish sawlog markets, with western and southern prices influenced by eastern regions. |
| Mei et al. (2010) | Southern U.S. | Sawtimber stumpage | ARIMA, VAR & EC | 128, quarterly | Identifies cointegration in 7 of 12 southern markets; multivariate models outperform univariate models in forecast accuracy. |
| Nanang (2000) | Canada | Regional softwood lumber prices | n/a (ADF test) | 68, quarterly | Detects cointegration among regional softwood lumber markets in Canada using ADF tests |
| Nanang (2009) | Ghana | Sawnwood, plywood and veneer exports | EC | 45, annually | Finds that three national policy initiatives reduced sawnwood exports but increased exports of plywood and veneer |
| Parajuli & Chang (2015) | Louisiana (U.S.) | Stumpage supply and demand | VEC | 59, annually | Shows that stumpage supply and demand are price inelastic in the long-run, using VEC modeling. |
| Parajuli et al. (2016) | Louisiana (U.S.) | Stumpage supply and demand | VEC | 59, annually | Confirms that VEC estimates of supply and demand elasticities are comparable to those from simultaneous equation modeling. |
| Song et al. (2011) | U.S. & Canada | Softwood | EC | 200, monthly | Finds Canadian softwood lumber supply is more price elastic than U.S. supply; U.S. tariffs have limited impact on imports. |
| Song et al. (2012) | U.S. | Wood energy | EC | 42, annually | Attributes decline in U.S. wood energy consumption to technological advancement and increased availability of substitutes. |
| Stordal & Nyrud (2003) | Norway | Sawlog | VEC | 60, monthly | Identifies cross-border price linkages, with Norwegian sawlog prices influenced by Swedish market prices. |
| Toppinen (1998) | Finland | Sawlog | EC | 144, monthly | Finds stumpage prices positively influence sawlog supply in both short and long run; demand also responsive to long run prices. |

## 3. Methodology

This study employs a VEC model to forecast key macroeconomic and industry-specific variables for U.S. states where the forestry and logging industry (NAICS 113) and its related sectors – wood manufacturing (NAICS 321) and paper manufacturing (NAICS 322) play a significant economic role. The primary objective of this analysis is to examine how these variables are expected to evolve in states where these industries are particularly important, as determined by their LQ value. Figure 1 below illustrates the modeling with a simple workflow. The remainder of this section is structured as follows: Section 3.1 formally defines the LQ value and its role in identifying industry-dependent states. Section 3.2 describes the data sources and variables included in the VEC model. Section 3.3 presents the theoretical foundation of the model, while Section 3.4 details how the empirical model is specified and estimated based on the data.

### *3.1. Location Quotients for Forest Industries*

To assess the relative importance of NAICS 113, 321 and 322 across U.S. states, we estimate state-level LQs. LQs are a standard measure used to compare the concentration of an industry within a specific region to its concentration at the national level, Equation 1 below shows its calculation.

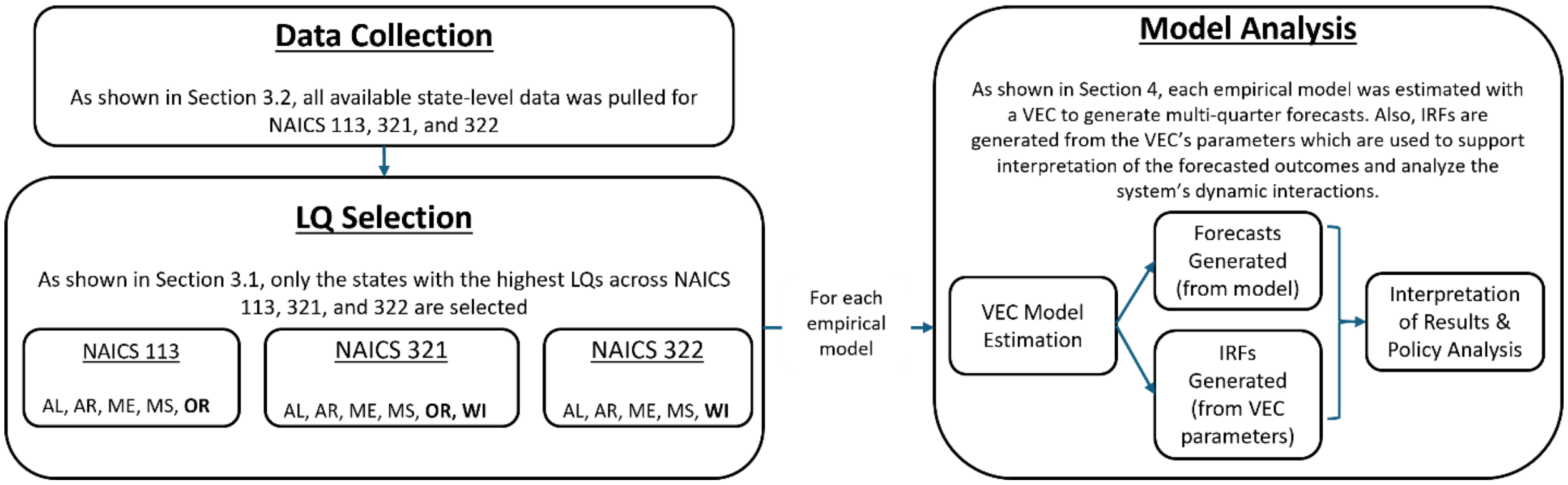

**Figure 1.** Modeling Workflow

$$LQ = \frac{I_r/E_r}{I_n/E_n} \tag{1}$$

To calculate LQ the following elements are used: $I_r$ is the industry in a region, $I_n$ is the industry in the nation, $E_r$ is employment in the region, and $E_n$ is employment in the nation. This metric calculates an industry's regional employment share in a given region relative to its share nationally (BLS, 2011). An industry is considered regionally significant if its LQ exceeds one, indicating a higher than average concentration of employment compared to the national level. While several states exhibit LQs above one, this study focuses on states where the LQs are consistently the highest across all 3 NAICS industries. The criteria of which states are picked will be detailed in Section 3.4.

### *3.2. Data*

To analyze state-level impacts in the U.S. forest sector, we use data from NAICS 113, 321 and 322. This dataset consists of quarterly macroeconomic data spanning from 2001Q1 to 2018Q4, with variable descriptions and sources summarized in Table 2. Recent data beyond 2018 was excluded following the forecasting literature to avoid distortions caused by the non-systemic shocks of the COVID-19 pandemic (Eicher et al., 2019; Bobeica & Hartwig, 2023; Schorfheide & Song, 2024). The variables included in the analysis represent key supply and demand factors: price and employment impact demand while output, wages, number of firms and price impact supply. To measure wages, total wages were used instead of wages per capita, as the latter assumes uniform earnings across workers – an overly restrictive assumption.

State-level price data was unavailable for each industry, so we used industry-level U.S. price indices as proxies for each NAICS code. This cannot capture variations in regional and/or state prices, but the U.S. price index should still be used rather than omitted to control for a crucial factor that can impact both supply and demand. While this substitution will likely introduce some measurement error, especially in regions where input costs or product markets diverge substantially from national averages, previous research has found that U.S. level forestry price indices generally track regional trends, especially over longer time horizons (Stordal & Nyrud, 2003). Therefore, any bias introduced is likely to attenuate rather than reverse observed trends. Similarly, quarterly output was proxied using annual state-level real gross domestic product (RGDP) in conjunction with quarterly U.S. RGDP. This adjustment follows a standard practice of distributing annual data into quarterly observations using national trends as a reference. To achieve this, state-level annual RGDP is converted to quarterly data by setting Q1 to one quarter of the annual RGDP, and the remaining quarters are proxied by following percent changes of U.S. Gross Output (GO)[1]. A key assumption in this process is that changes in state-level GDP are primarily driven by NAICS 113, 321, and 322, rather than other sectors. Related literature has also faced these data limitations and used similar approaches to maximize the use of available data (Stordal & Nyrud, 2003; Daigneault et al., 2016). Appendix A provides summary statistics for all variables across the three NAICS industries. No missing observations were identified, and all variables exhibit reasonable ranges, standard deviations, and trends. Although wages appear highly variable, they primarily follow a cyclical pattern. When accounting for this cyclicality, a gradual linear trend emerges over time, as discussed in the results section.

**Table 2**. Variable List. Includes variable names, descriptions, and sources used to model NAICS 113, 321 and 322 at the state-level.

| Variable | Description | Years* | Source |
|---|---|---|---|
| Employment | State-level employment for NAICS 113, 321 & 322 (in thousands, not seasonally adjusted) | 2001-2018 | BLS |
| Number of Firms | State-level employment for private NAICS 113, 321 & 322 (not seasonally adjusted) | 2001-2018 | BLS |
| Price | Producer Price Index for NAICS 113, 321, & 322 (in 2018 dollars, not seasonally adjusted) | 2001-2018 | BLS |
| Wages | State-level total nominal wage for private NAICS 113, 321 & 322 (in thousands, not seasonally adjusted) | 2001-2018 | BLS |
| Output | Proxy of state-level RGDP (in millions, not seasonally adjusted) | 2001-2018 | BEA |
| *Due to data availability, NAICS 321 & 322 only span 2004-2018 | | | |

### *3.3. Theoretical Model*

To analyze the data, this section reviews the framework of cointegrated models, explains the interpretation of their estimates, and discusses how these estimates are used for forecasting. Non-stationary time-series models have widely used in the literature, particularly in studies of U.S. and Canadian lumber markets, which are typically non-stationary (Nanang, 2000; Stordal & Nyrud, 2003; Yin & Baek, 2005; Song et al., 2011; La & Mei, 2015; Gendron et al., 2019). In the presence of cointegration, Equation 2 is employed to run an EC model which captures short-run adjustments toward a long-run equilibrium (Lambert, 2013).

[1] This is for data starting 2005Q1 onward, 2001Q1 to 2004Q4 required different imputation due to a lack of quarterly U.S. GO data. To fill this gap, we imputed by following the average trend of proxied quarterly state-level RGDP and lining up values with annual state-level RGDP.

$$EC: \Delta y_t = \delta_0 + \sum_{i=1}^{p} \delta_i \Delta x_{t-i} + \sum_{j=1}^{k} \mu_i \Delta y_{t-j} - \lambda(y_{t-1} - \alpha - \beta x_{t-1}) + \epsilon_t \quad (2)$$

When calculating the EC model in the above equation, the term $(y_{t-1} - \alpha - \beta x_{t-1})$ represents the cointegrating relationship between $x_t$ and $y_t$. Even if the linear combination $(y_t - \beta x_t)$ is stationary, but $y_t$ and $x_t$ are nonstationary, then $y_t$ and $x_t$ are a cointegrated series (Yin & Xu, 2003). Note that $\lambda$ is the error correction term, which determines how quickly deviations from equilibrium are corrected. When there are multiple cointegrating equations, a vector of EC models known as a VEC model should be used (Nanang, 2000; Stordal & Nyrud, 2003; Parajuli & Chang, 2015; Gendron et al., 2019), as shown in Equation 3.

$$VEC: \Delta X_t = \sum_{i=1}^{k-1} \Gamma_i \Delta X_{t-i} + \Pi X_{t-1} + \mu + \phi D_t + \epsilon_t \quad (3)$$

To calculate the VEC model in the above equation, $X_t$ is a vector of variables, $k$ is the lag length, $\mu$ are constant terms, $\phi D_t$ measures the number of cointegrating equations, $\Pi \Delta X_{t-i}$ and $\Gamma X_{t-1}$ represent the lag structure, $\Pi$ are coefficients for the long-run equilibrium relationships, and $\Gamma$ are coefficients for the short-run dynamics.

We use a VEC model because it allows us to measure bi-directional and cross directional relationships between variables (unlike standard regression models, which assume a single-directional relationship). In addition, a VEC model allows us to handle multiple cointegrated equations, which are essential given the interconnected nature of supply and demand in the forest sector. By utilizing a VEC model, we can simultaneously estimate long-run equilibrium relationships and short-run adjustments for key macroeconomic variables. Our model primarily focuses on employment, output and wages, capturing both supply and demand dynamics in the forest sector. These estimated relationships are then used for forecasting, an essential tool for projecting future trends in forest related industries (Pattanayak et al., 2002; Xu et al., 2004; Baek, 2012). Ultimately, the VEC model enables us to provide a comprehensive empirical framework for understanding the interdependencies within the forest sector and forecasting its future economic trajectory (Toppinen, 1998; Toppinen & Kuuluvainen, 2010; Baek, 2012; Parajuli & Chang, 2015; Daigneault et al., 2016; Parajuli et al., 2016).

### *3.4. Empirical Model*

Building on the framework of Gendron et al. (2019), we analyze five key variables – employment, output, wages, number of firms, and price – as a system of equations within a five-dimensional VEC model. This approach allows us to capture the interdependence of these variables within the forest sector at the state-level. The general form of our empirical model is shown in Equation 4.

$$\Delta X_t = \sum_{i=1}^{k-1} \Gamma_i \Delta X_{t-i} + \Pi X_{t-1} + \mu + \phi D_t + \epsilon_t \quad (4)$$
$$X_t = [output_t, employment_t, wage_t, numberFirms_t, price_t]'$$
$$t = 2001Q1 - 2018Q4$$

The equation above includes the following elements: $X_t$ is a vector of our variables, $k$ is the lag length, $\mu$ is a vector of constant terms, $\phi D_t$ is a vector that measures the number of cointegrating equations (aka rank), $\Pi \Delta X_{t-i}$ and $\Gamma X_{t-1}$ are vectors that represent the lag structure, $\Pi$ are coefficients for long-run relationships, and $\Gamma$ are coefficients for short-run dynamics. The $\mu$ is used for the Johansen restriction, and $\Delta X_t$ estimates measure the impact of everything used to analyze the impact of one to four variables (depending on the lag structure of each model). It is important to note that there are 16 empirical models built: NAICS 113 includes 5 states, NAICS 321 includes 6 states, and NAICS 322 includes 5 states. These numbers differ per sector since only four out of the six states have high LQ values for all three sectors. The criteria for selecting each state in each industry were the highest LQ values consistently across NAICS 113, 321 and 322. Based on this criterion, we identify six states as the most forest-dependent: Alabama, Arkansas, Maine, Mississippi, Oregon, and Wisconsin (see Appendix B for details)[2].

To determine the appropriate model specifications for each NAICS code and state, we applied Johansen's cointegration test and lag selection criteria. The rank of cointegration and lag length vary by model, but each was selected based on statistical criteria to ensure optimal performance. Given our sample size, we used the following lag selection tests: Akaike Information Criterion (AIC), Final Prediction Error (FPE), and sequential likelihood-ratio (LR) (Liew, 2004). Whenever there were conflicts in lag selection tests, only one out of the three tests would have a different lag, so we would use the optimal lag. If all three conflicted in future study, then further care would be needed in choosing the optimal lag, such as testing the performance of the model with all potential lags. Each model was further tested for autocorrelation using the Lagrange-multiplier (LM) test, which confirmed that residuals were not autocorrelated in each model and their selected lags. Additionally, we tested model stability across time and found that the cointegration relationships and short-run dynamics in all 16 empirical models are stable. This suggests that trends are robust to variation within 2001Q1-2018Q4, please see Appendix A for unit root graphs with respective eigenvalues. To assess normality, we conducted the Jarque-Bera, Kurtosis and Skewness tests. The test results indicated that for NAICS 113 and 321 two out of the five equations exhibited non-normal residuals, while for NAICS 322 one out of the five equations exhibited non-normal residuals. However, as Johansen (1995) notes, mild departures from normality do not invalidate inference as long as key conditions – autocorrelation, stationarity, structural breaks, and sample size considerations – are properly addressed, which is the case for our study (please see Appendix C for full diagnostic results[3]).

[2] Wisconsin's LQ is not significant for NAICS 113, but extremely high for NAICS 321 and 322. Oregon's LQ is not significant for NAICS 322, but extremely high for NAICS 113 and 321

[3] Details for the other tests (Johansen Tests for Cointegration and the Lagrange Multiplier Test) are available upon request.

To analyze dynamic interactions among variables, we employ impulse response functions (IRFs), a common tool in macroeconomic and VEC modeling. IRFs trace the effect of a one-time shock to one variable on the future values of all variables in the system, allowing us to distinguish both short-run adjustments and long-run responses. Following Gendron et al. (2019), we use these IRFs to generate out-of-sample forecasts based on how the system responds to shocks over time. To illustrate how this works in practice, consider a one-time shock to employment. The example IRF plot for Maine employment shown in Figure 2 simulates how this employment shock propagates through the rest of the system. For instance, wages experience an immediate spike that then levels off after two quarters. Interestingly, we also observe a seasonal oscillation: each year, wages tend to spike in the first quarter and level off by the fourth. These patterns emerge naturally from the estimated relationships among variables in the VEC model and reflect the historical dynamics captured in the data. These detailed dynamics provide insight into how one variable affects the others over time, and this is a key strength of VEC models. IRFs not only help us interpret short and long run interdependencies but also provide a foundation for generating realistic shock-based forecasts.

This approach mirrors methodologies used in the lumber forecasting literature, where VEC models have been applied to project supply and demand dynamics across regions and countries. By integrating VEC based forecasting and IRFs, this empirical strategy provides a comprehensive understanding of the macroeconomic forces shaping the forest sector, offering valuable insights for policymakers and industry leaders. These insights can then be used for best informed policy intervention to mitigate the damage of economic shifts and supply support to its affected communities.

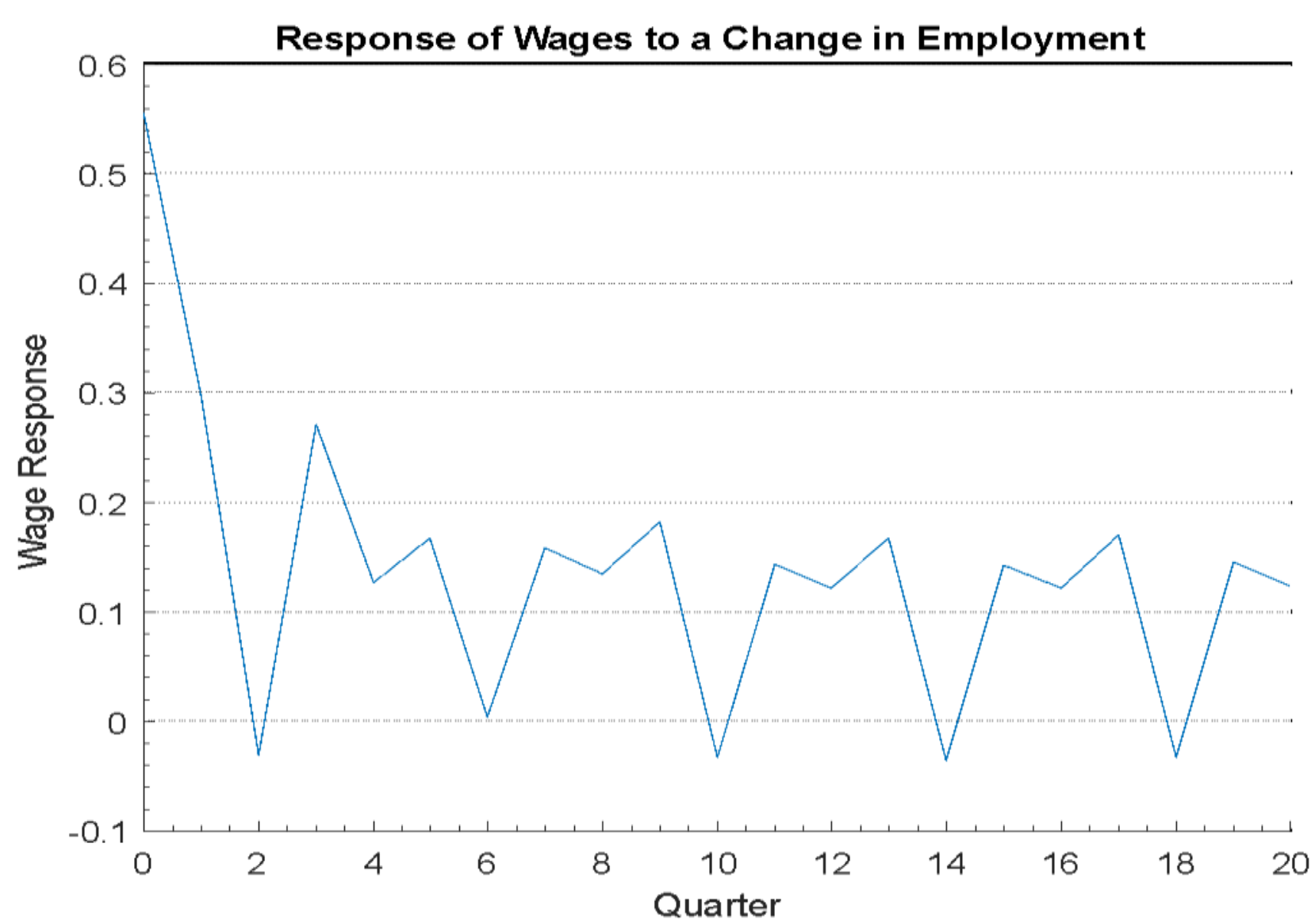

**Figure 2.** IRF for Maine Employment in NAICS 113. Orthogonalized IRF generated to support the response of wages to a shock in employment in Maine.

## 4. Results and Discussion

We forecasted trends for the 5 empirical models of NAICS 113 (forestry and logging), the 6 empirical models of NAICS 321 (wood product manufacturing), and the 5 empirical models of NAICS 322 (paper manufacturing) and compared them to historical data to assess forecast accuracy[4]. The forecasting horizon spans five years (2018Q1 to 2023Q1), using data from 2001Q1-2018Q1 for model estimation. Forecasts are compared to historical trends to assess accuracy and highlight implications for regional policy and industry sustainability. Importantly, these forecasts do not account for the COVID-19 pandemic, as it represents a non-systemic shock that would distort long-run industry trends. As highlighted in the methodology within Section 3, wages in the forestry sector exhibit a cyclical pattern. After adjusting for this cyclical behavior, only a modest upward trend persists.

Our study focuses on the six states where LQs are consistently the highest across all 3 NAICS industries (forestry, wood and paper manufacturing): Alabama, Arkansas, Maine, Mississippi, Oregon, and Wisconsin (see Table 3). These states represent a diverse geographic range, covering the Northeast, North Central, South Central, and Pacific West regions. The high economic dependence of communities in these states on the forestry sector makes them particularly vulnerable to structural changes, highlighting the importance of regional policy interventions for their economic stability and industry sustainability. Our vector error correction (VEC) model proves effective in capturing long-run dynamics in employment, output, wages, and firm count, offering valuable insights for policymakers and stakeholders seeking to improve regional resilience, develop workforce transition strategies, or prioritize infrastructure investment. The following subsections summarize key industry-level forecasts and discuss implications for targeted policy intervention: Section 4.1 presents and discusses the results for NAICS 113, followed by Section 4.2 for NAICS 321, and Section 4.3 for NAICS 322. Section 4.4 outlines both the limitations and advantages of the empirical framework used in these forecasts, and Section 4.5 provides some suggestions for future study. Lastly, Section 4.6 provides some suggestions for policy-makers of the states we empirically modeled.

### *4.1 NAICS 113 (Forestry and Logging)*

Forecasts for NAICS 113 shown in Figure 3-4 indicate a general decline in employment and the number of firms, while output remains stable and wages show a modest upward trend. Maine is the only state where employment exhibits seasonal patterns. Oregon is a notable outlier with significantly higher levels of employment, output, and wages, although the direction and rate of change are consistent with the broader group. The contraction in employment and business activity pose a serious challenge to communities with tax bases and household incomes dependent on this sector. However, the observed stability in output and wage growth suggests that the remaining firms are either more productive or benefit more from improved pricing or efficiency. Policymakers and industry leaders may interpret these findings as a call to action: targeted policy intervention and/or economic diversification efforts may be required in communities where mill closures and labor market contraction are likely, while those maintaining output may benefit from workforce retraining programs to transition displaced workers into higher-value roles within the sector.

[4] Details on empirical model specification were driven by LQ values, see Section 3.4 for further details.

**Table 3.** LQ Summary Statistics: This table shows summary statistics of the LQ value per state and per region spanning 2004Q1-2018Q4[5]. Bolded are the states we selected to empirically model, in parentheses are low LQ values (i.e. less than one). Note that "–" means we didn't have data, and the bottom half of each region are states where the LQ values are low across all 3 sectors (NAICS 113, 321 & 322).

| | | LQ Summary Statistics | | | | | | | | |
|---|---|---|---|---|---|---|---|---|---|---|
| | | NAICS 113 | | | NAICS 321 | | | NAICS 322 | | |
| **Region** | **State** | **Mean** | **Min** | **Max** | **Mean** | **Min** | **Max** | **Mean** | **Min** | **Max** |
| Pacific West | **OR** | **8.69** | **7.65** | **9.68** | **4.72** | **4.52** | **4.88** | 1.02 | 0.85 | 1.15 |
| | WA | 3.49 | 2.98 | 4.31 | 1.62 | 1.39 | 1.75 | 1.06 | 0.91 | 1.30 |
| | AK | 1.87 | 0.43 | 5.81 | - | - | - | - | - | - |
| | *(CA)* | - | - | - | *(0.55* | *0.50* | *0.65* | *0.51* | *0.49* | *0.53)* |
| South Central | **AL** | **5.52** | **5.13** | **5.97** | **2.75** | **2.51** | **3.13** | **2.18** | **2.03** | **2.35** |
| | **AR** | **4.97** | **4.38** | **5.79** | **2.91** | **2.73** | **3.13** | **2.97** | **2.80** | **3.14** |
| | KY | - | - | - | 1.95 | 1.74 | 2.09 | 1.72 | 1.58 | 1.83 |
| | LA | 3.54 | 3.08 | 3.89 | 1.19 | 1.03 | 1.28 | 1.35 | 1.20 | 1.55 |
| | **MS** | **7.00** | **6.52** | **7.59** | **3.10** | **2.81** | **3.44** | **1.30** | **1.16** | **1.47** |
| | *(TN)* | *(0.72* | *0.59* | *0.96)* | 1.55 | 1.47 | 1.63 | 1.89 | 1.59 | 2.00 |
| | *(OK)* | - | - | - | *(0.52* | *0.38* | *0.67* | *0.62* | *0.47* | *0.77)* |
| | *(TX)* | - | - | - | *(0.70* | *0.67* | *0.75* | *0.57* | *0.53* | *0.65)* |
| South East | GA | 2.98 | 2.61 | 3.30 | - | - | - | - | - | - |
| | NC | 1.82 | 1.68 | 2.01 | - | - | - | - | - | - |
| | SC | 3.85 | 3.60 | 4.22 | - | - | - | - | - | - |
| | VA | 1.28 | 1.11 | 1.58 | - | - | - | - | - | - |
| North East | **ME** | **9.06** | **5.94** | **11.18** | **2.67** | **2.36** | **2.94** | **3.87** | **2.76** | **4.45** |
| | PA | - | - | - | 1.33 | 1.26 | 1.42 | 1.43 | 1.29 | 1.53 |
| | VT | - | - | - | 2.09 | 1.78 | 2.65 | 1.06 | 0.81 | 1.26 |
| | *(NH)* | 1.59 | 1.20 | 1.89 | 1.00 | 0.89 | 1.16 | *(0.76* | *0.49* | *1.30)* |
| | *(WV)* | 3.10 | 2.36 | 3.84 | 2.71 | 2.42 | 2.94 | *(0.29* | *0.26* | *0.32)* |
| | *(CT)* | - | - | - | *(0.26* | *0.21* | *0.30* | *0.79* | *0.69* | *0.95)* |
| | *(MA)* | - | - | - | *(0.26* | *0.23* | *0.27* | *0.99* | *0.86* | *1.19)* |
| | *(MD)* | - | - | - | *(0.32* | *0.28* | *0.37* | *0.46* | *0.38* | *0.59)* |
| | *(NJ)* | - | - | - | *(0.24* | *0.20* | *0.31* | *0.99* | *0.93* | *1.06)* |
| | *(NY)* | - | - | - | *(0.30* | *0.27* | *0.34* | *0.66* | *0.61* | *0.69)* |
| | *(OH)* | - | - | - | *(0.82* | *0.73* | *0.89)* | 1.35 | 1.27 | 1.47 |
| North Central | IN | - | - | - | 1.62 | 1.48 | 1.71 | 1.19 | 1.05 | 1.31 |
| | MN | - | - | - | 1.44 | 1.33 | 1.56 | 1.29 | 1.22 | 1.41 |
| | **WI** | - | - | - | **2.25** | **2.16** | **2.35** | **3.86** | **3.61** | **4.09** |
| | *(IA)* | - | - | - | 2.09 | 1.94 | 2.28 | *(0.88* | *0.82* | *0.95)* |
| | *(IL)* | - | - | - | *(0.39* | *0.36* | *0.42)* | 1.23 | 1.16 | 1.28 |
| | *(MI)* | *(0.89* | *0.66* | *1.21* | *0.75* | *0.62* | *0.86* | *0.99* | *0.88* | *1.13)* |
| | *(MO)* | - | - | - | *(0.99* | *0.87* | *1.10* | *0.95* | *0.91* | *1.00)* |

[5] Note that for NAICS 113 the timespan is 2001Q1-2018Q4.

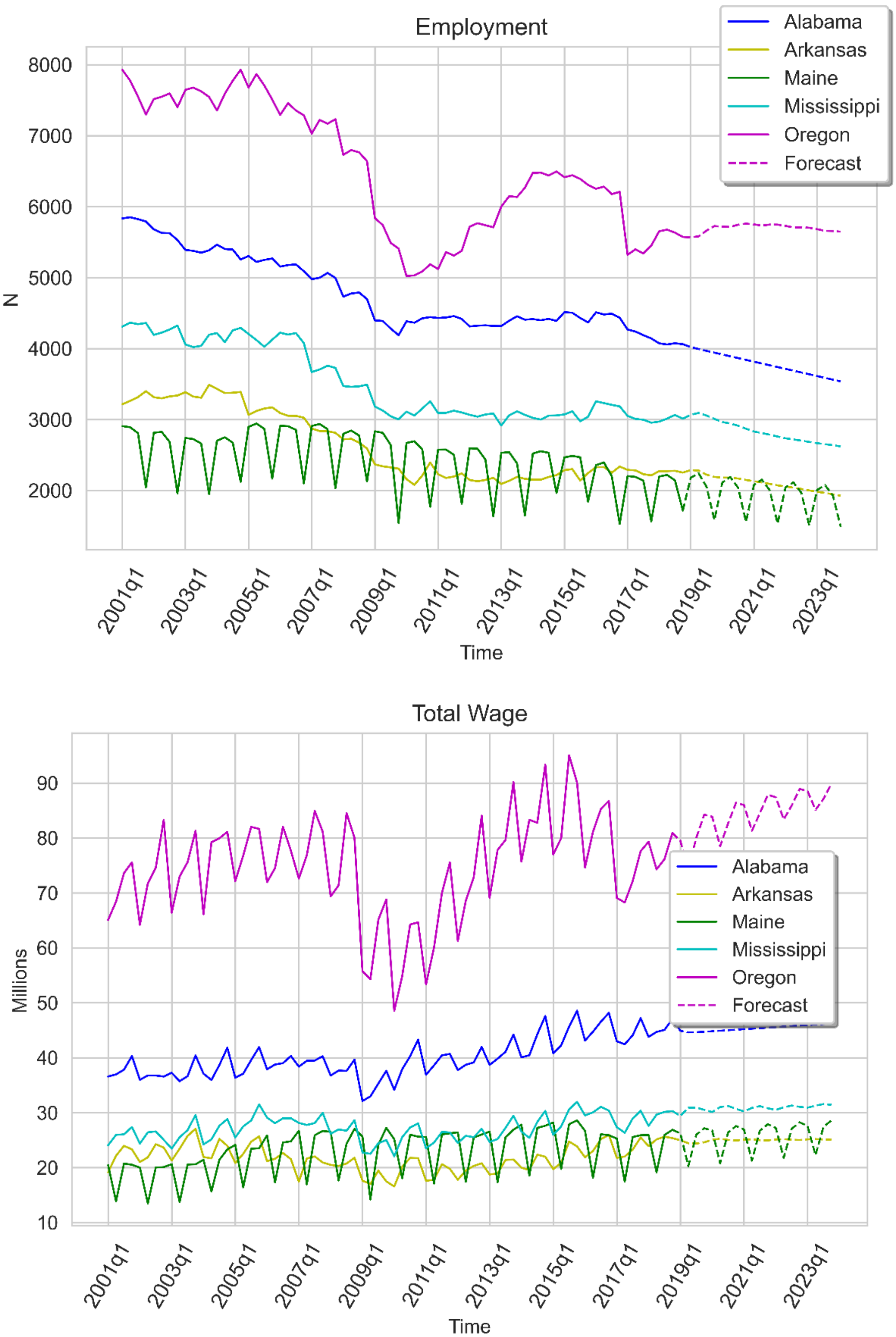

**Figure 3.** Key state NAICS 113 variable forecasts for employment and total wage (out of sample forecasts projected for five years assuming the non-systematic shock of the Coronavirus pandemic did not occur).

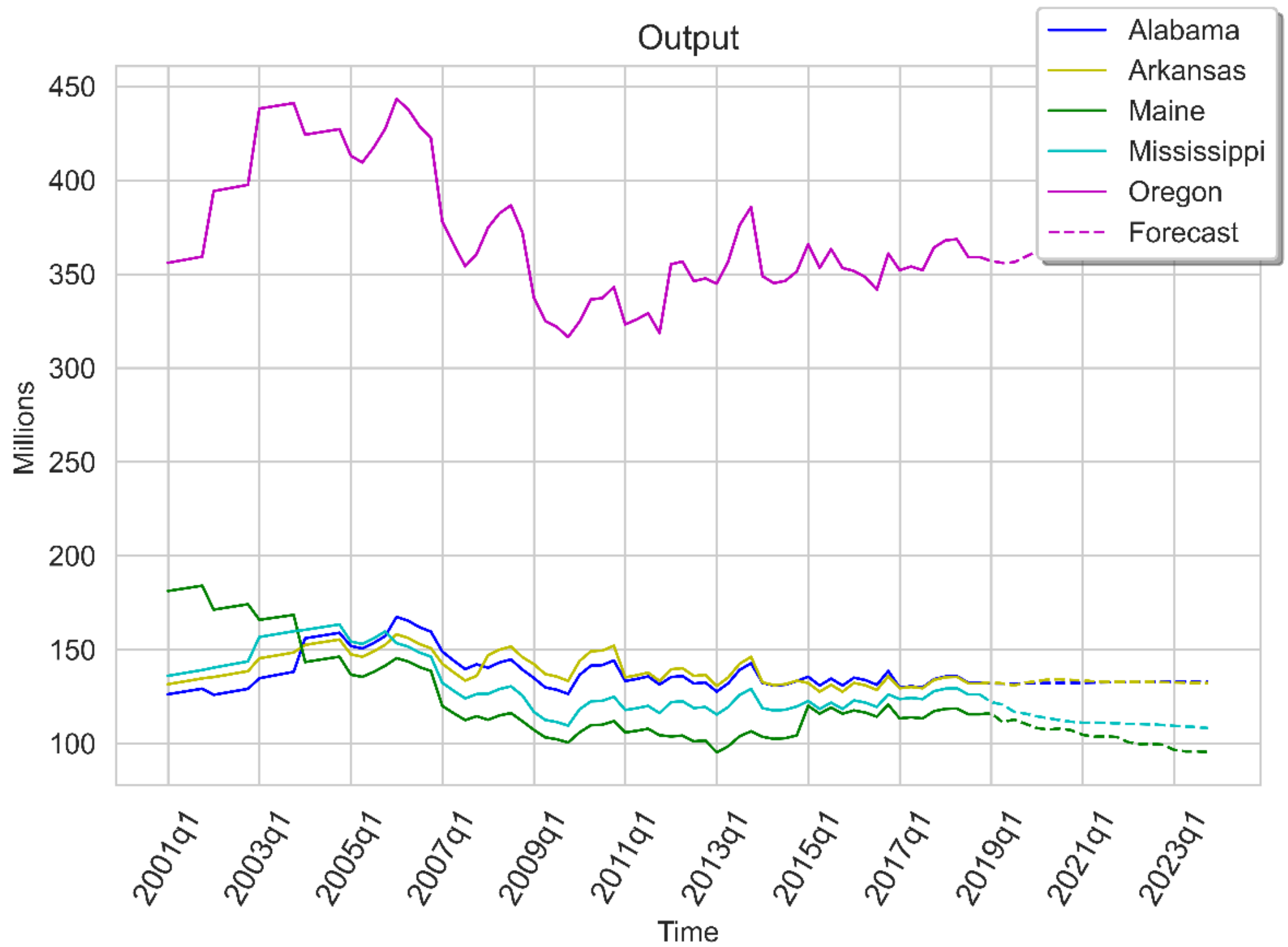

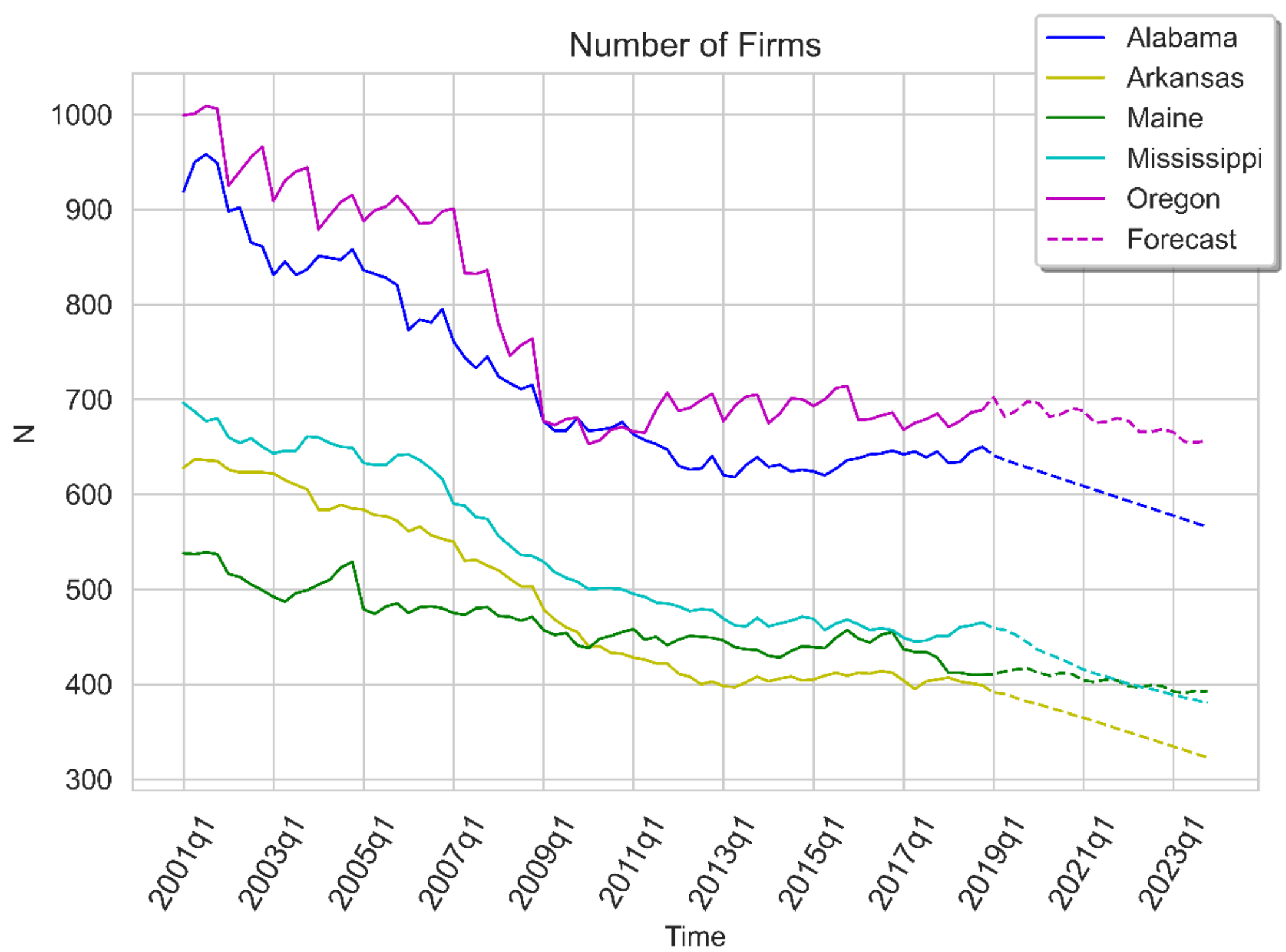

**Figure 4.** Key state NAICS 113 variable forecasts for output and number of firms (out of sample forecasts projected for five years assuming the non-systematic shock of the Coronavirus pandemic did not occur).

### *4.2 NAICS 321 (Wood Product Manufacturing)*

The results for NAICS 321 as shown in Figure 5-7 provide various insights. Employment and wages fell sharply during the financial crisis (2008-2010) but rebounded gradually over time in most states. Post-crisis employment continued to rise steadily in Alabama and Oregon, while wages in Alabama increased at a faster rate compared to the other states in this analysis. Firm count has been steadily declining for most states with the exception of Alabama and Oregon whom have instead leveled off. Output followed a similar trajectory – dropping during the financial crisis, recovering to pre-crisis levels over time, and then stabilizing. Arkansas saw particularly volatile output growth between 2012 and 2015, while Alabama, Mississippi, and Wisconsin experience similar but more moderate output surges from 2014 to 2018.

In summary, NAICS 321 exhibits more favorable labor market dynamics than NAICS 113 does. The upward trend in employment is a promising sign for local economic development, suggesting that targeted support to sustain firm viability – such as infrastructure improvements or capital investment programs – may yield significant returns. However persistent declines in the number of firms in some regions signal underlying structural challenges. As such, policy measures that reduce entry barriers, incentivize mill modernization, or provide technical assistance could play a key role in supporting not only stability, but also industry longevity, and employment growth.

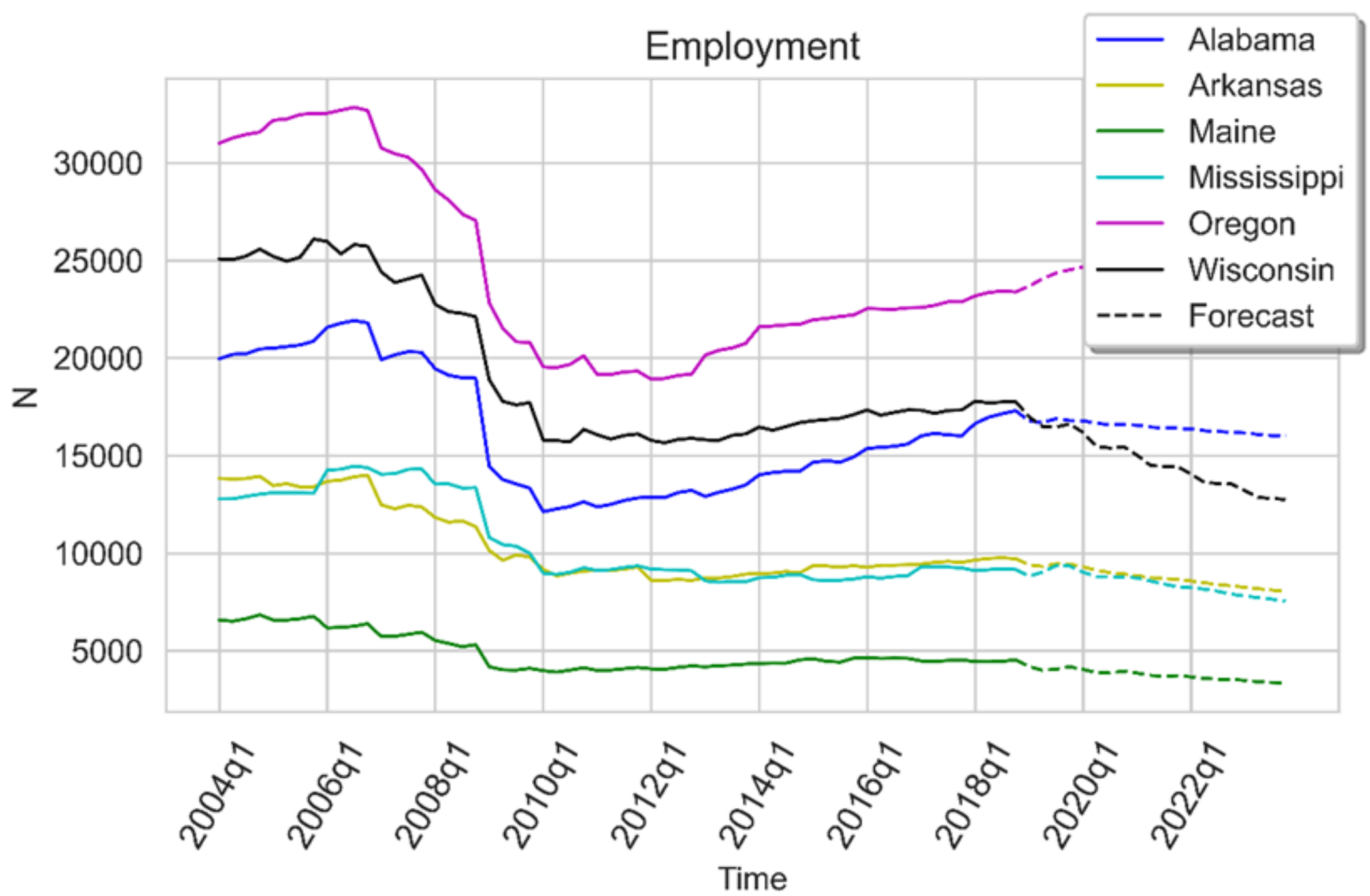

**Figure 5.** Key state NAICS 321 variable forecasts for employment (out of sample forecasts projected for five years assuming the non-systematic shock of the Coronavirus pandemic did not occur).

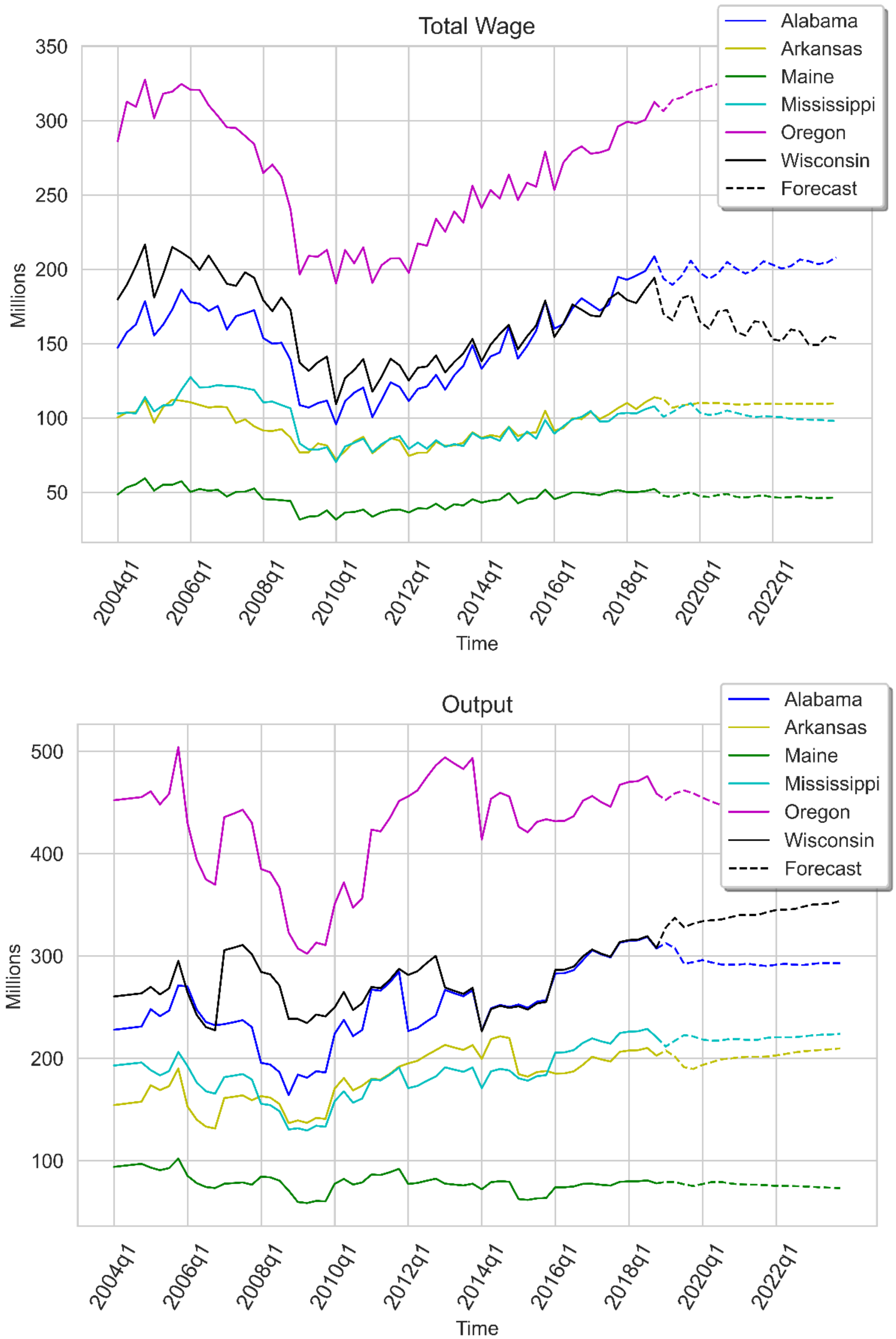

**Figure 6.** Key state NAICS 321 variable forecasts for total wages and output (out of sample forecasts projected for five years assuming the non-systematic shock of the Coronavirus pandemic did not occur).

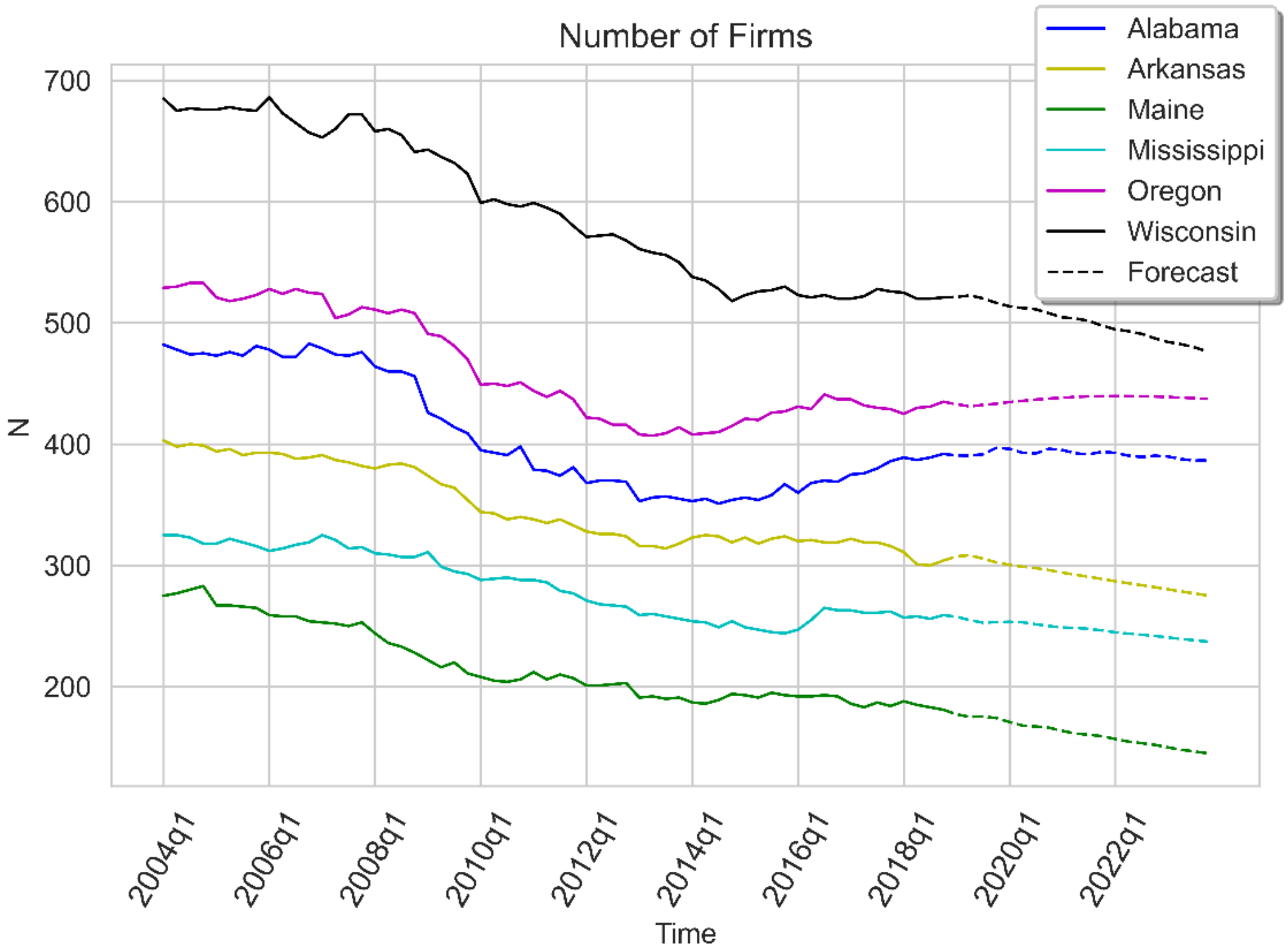

**Figure 7.** Key state NAICS 321 variable forecasts for number of firms (out of sample forecasts projected for five years assuming the non-systematic shock of the Coronavirus pandemic did not occur).

### *4.3 NAICS 322 (Paper Manufacturing)*

The results for NAICS 322 shown in Figures 8-9 have a concerning trajectory. Although wages remain relatively stable with seasonal fluctuations, employment, output, and the number of firms are all forecasted to decline. Maine shows a particularly steep decline across all indicators, suggesting that this sector is under acute strain. Wisconsin, by contrast, has much higher levels across all indicators, but shares a similar rate of decline most of the other states in this analysis exhibit. A noteworthy exception is Arkansas in that its output does not decrease like the other states but instead remains level. These results highlight the urgency for policy support in regions where the paper manufacturing industry remains a major employer. In particular, Maine's sharp contradictions could result in substantial economic dislocation unless mitigated by targeted investment or retraining programs. The stability in wages – despite firm and job losses – suggests that remaining jobs may be higher-skilled or more capital-intensive, reinforcing the need for human capital development strategies. Additionally, the relatively resilient performance of Wisconsin suggests a potential case study for how public and private sector collaboration can sustain a mature industry in transition.

### *4.4. Model Advantages and Limitations*

Our use of the VEC model reflects both data constraints and the need to accurately model long-run relationships across interdependent variables. Literature on U.S. lumber markets consistently supports the use of cointegrated models (Nanang, 2000; Nyrud, 2003; Yin & Baek, 2005; Song et al., 2011; Gendron et al., 2019). VEC model also enables the study of multiple bidirectional relationships unlike constrained EC models – essential for understanding how changes in employment, wages, and output interact over time. Forecasts were validated by comparing predicted values with actual data for 2016-2018 (see Figures A1-A3). Aside from Oregon, which had a larger degree of volatility, the forecasts aligned closely with observed trends, affirming the models sustainability for regional industry forecasting. However, some limitations should be noted. First, state-level RGDP data is only reported annually, requiring us to proxy quarterly output using U.S. RGDP (Mei et al., 2010). This trade-off allowed for more robust forecasting but introduces some aggregation error. Price data was another constraint, so we used the 2017 PPI for lumber and wood products, as done by Baek (2012) and Nanang (2000) in the absence of more granular price data. Moreover, there was no data available for capital, meaning we could not account for how capital investments might affect mill operations and log demand. This omission is important since it misses the key balance between labor and capital, and how improvements in capital via automation and efficiency significantly changed demand. Further research incorporating capital data would be highly valuable. Furthermore, our analysis focused on six states; future research should examine other forest-intensive regions such as Kentucky, Louisiana, and Tennessee since they have consistently high LQ values for all three industries.

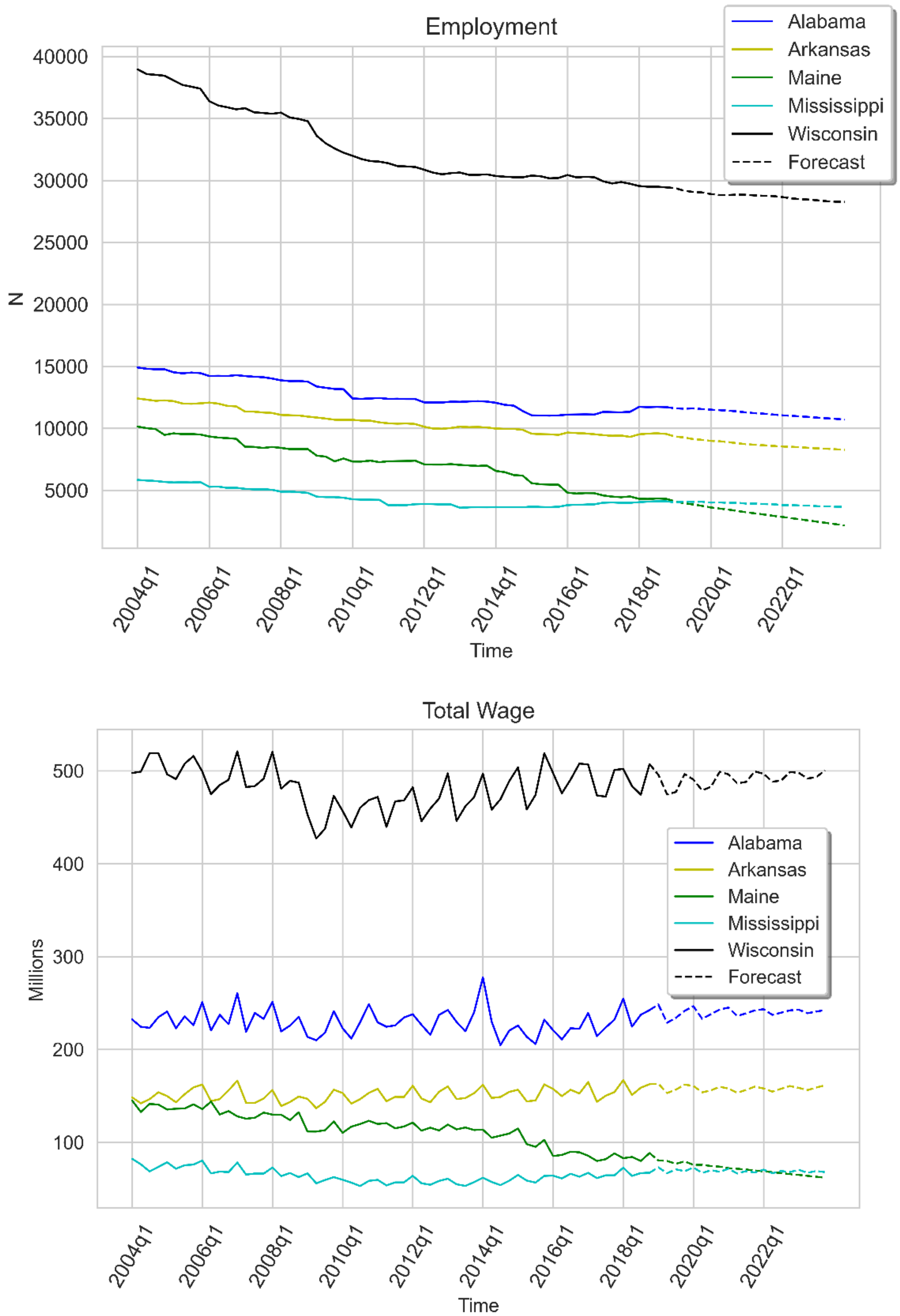

**Figure 8.** Key state NAICS 322 variable forecasts for employment and total wage (out of sample forecasts projected for five years assuming the non-systematic shock of the Coronavirus pandemic did not occur).

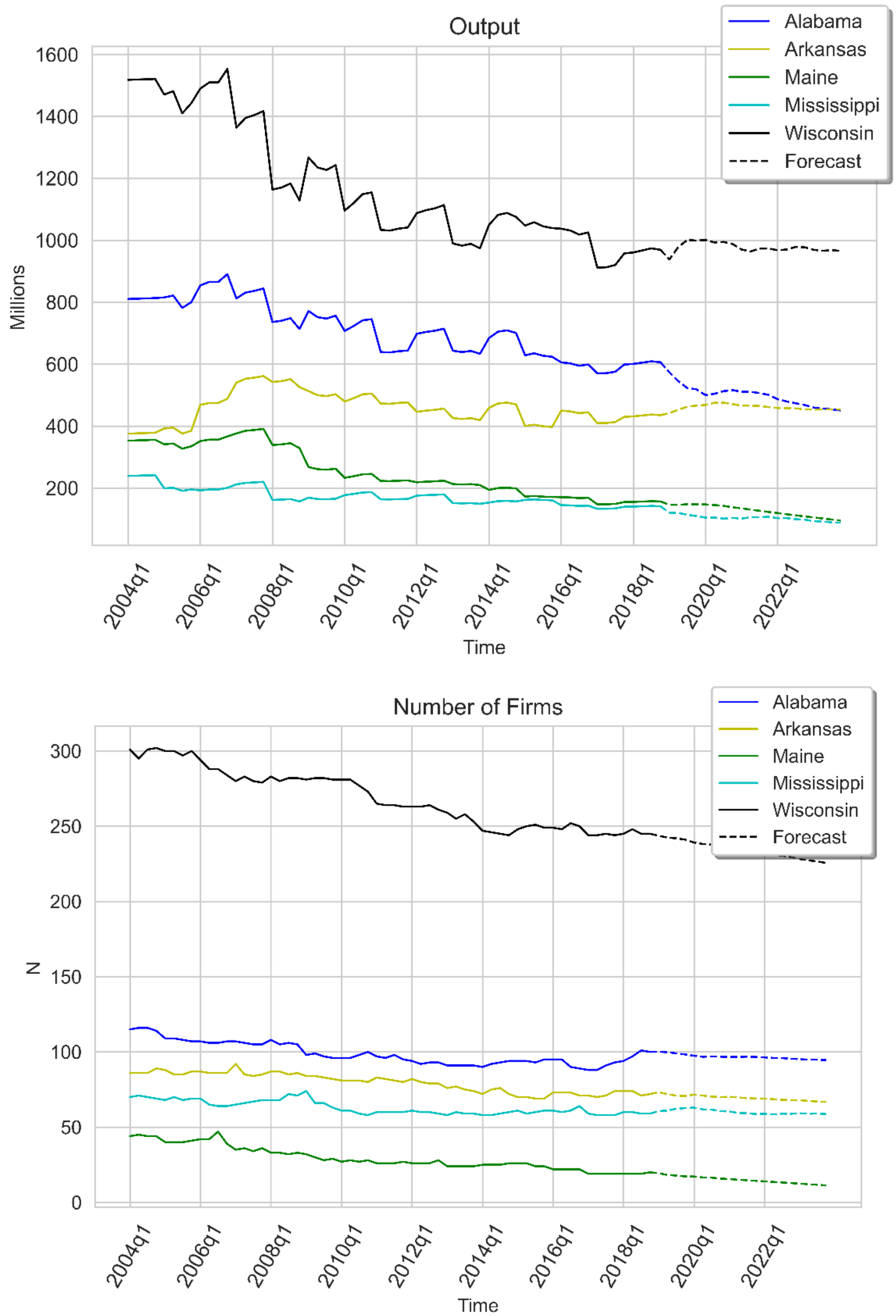

**Figure 9.** Key state NAICS 322 variable forecasts for output and number of firms (out of sample forecasts projected for five years assuming the non-systematic shock of the Coronavirus pandemic did not occur).

A second limitation is the sample size of the available data: 72 observations for NAICS 113 and 60 observations for NAICS 321 and 322. While these sample sizes are comparable to previous studies, additional data would enhance statistical power robustness, accordingly improving the reliability of forecasts (Toppinen, 1998; Malaty et al., 2006). The literature does not have a consensus on the ideal sample size for econometric forecasting, but related studies often work with between 57 to 69 observations (Toppinen, 1998; Nanang, 2000; Stordal & Nyrud, 2003; Hetemaki et al., 2004; Parajuli & Chang, 2015; Parajuli et al., 2016; Gendron et al., 2019;). Some related studies have been conducted with less than five years of data (Toppinen, 1998; Stordal & Nyrud, 2003; Yin & Xu, 2003; Malaty et al., 2006).

### *4.5. Future Study*

While this study focused on six of the most forest-dependent U.S. states, we attempted to expand this analysis to a broader regional scope – covering approximately 15 states for NAICS 113, and 40 states for NAICS 321 and 322. However, the results were inconclusive, likely due to heterogeneity within each region, potential differences in data quality, and potential regional economic dynamics. Future research should aim to refine the structure of regional analysis beyond geographic groupings, this could include shared forest composition, mill capacity and/or labor force characteristics. This could be further enhanced with the use of spatial econometric techniques which could potentially capture spillover effects across neighboring states. These interlinked supply chains and/or shared labor markets are a crucial factor for policymakers to manage.

### *4.6. Suggested Policy Actions*

As we have illustrated, there are distinct differences at the state and/or industry level, so policy will need to be tailored to particular scenarios. Accordingly, we provide suggested policy actions per scenario:

<u>Scenario 1</u>: Output stable; employment and firm count declining; wages rising modestly (e.g., NAICS 113 – Forestry and Logging)

> When output remains steady despite declining employment and number of firms, policymakers should prioritize workforce retraining and upskilling programs. These initiatives can help transition displaced workers into higher-skill roles within or beyond the sector, improving labor market resilience without compromising productivity.

<u>Scenario 2</u>: Employment and wages increasing; firm count declining or leveling (e.g., NAICS 321 – Wood Product Manufacturing)

> In cases where labor market indicators are strong but the number of firms is shrinking, policymakers should consider lowering entry barriers, offering mill modernization grants, and/or providing technical assistance to encourage entrepreneurship and bolster firm survival. Supporting firm formation and continuity can help sustain the employment momentum already underway (and respectively wages).

<u>Scenario 3</u>: Wages stable; employment, output, and firm count declining (e.g., NAICS 322 – Paper Manufacturing)

> This scenario suggests structural contraction. In such cases, targeted support from policymakers could take two forms. First, targeted investment in at-risk firms could prevent further losses in firm counts, and respectively employment and output. Second, transition assistance and retraining programs could redeploy affected workers into other segments of the forest sector or adjacent industries, preserving community-level economic vitality.

## 5. Summary & Conclusions

During the past three decades, industries in the U.S. forest sector have faced substantial challenges due profound structural changes driven by mill closures, technological shifts, and changes in global demand (Toppinen & Kuuluvainen, 2010; Crandall et al., 2017; Woodall et al. 2011; Conrad et al., 2018; Hodges et al., 2011; Poudel & Dahal, 2025). These changes have disproportionately impacted rural communities where forestry-related industries serve as economic linchpins. This study provides region-specific forward-looking evidence to support more informed, proactive policymaking, particularly in rural areas where these industries are economic cornerstones. We forecast key macroeconomic trends for three core forest sector industries within the U.S.: forestry and logging (NAICS 113), wood manufacturing (NAICS 321) and paper manufacturing (NAICS 322). We focused on six key states as being the most forest-dependent in the country, Alabama, Arkansas, Maine, Mississippi, Oregon and Wisconsin – identified using a location quotient measure and supported in prior literature.

To capture both supply and demand dynamics, we modeled state-level industry variables including employment, wages, number of firms, output (proxied with state-level real gross domestic product), and prices for U.S. lumber and wood products (Andrei & Andrei, 2015). Using quarterly data from 2001Q1 to 2018Q4, we constructed 16 state-industry VEC models, chosen for their ability to estimate long-run equilibrium relationships in cointegrated time series data such as the U.S. forest sector (Nanang, 2000; Stordal & Nyrud, 2003; Yin & Baek, 2005; Song et al., 2011). While data constraints precluded the inclusion of stumpage prices or capital investment variables, the modeling framework selected remains one of the most appropriate for this sector. In addition to forecasting, the VEC model enables insight into cross-sector interactions, such as how labor market shifts may influence output or firm dynamics (Zhang & Xie, 2019).

Our forecasts indicate that the forestry and logging industry is likely to experience continued declines in both employment and firm counts, even as output remains steady and wages gradually rise. This suggests a form of industrial resilience that may not translate to labor market stability, thereby highlighting the need for targeted state-level policy responses to mitigate job losses and strengthen rural economies. Policymakers should anticipate the ripple effects of declining employment in mill regions and consider economic diversification strategies to stabilize community tax bases and employment. The wood product manufacturing industry exhibits similar trends, but with a more optimistic outlook: employment is expected to rise. This trend suggests a comparatively resilient subsector where less intervention may be needed, allowing policymakers to reallocate limited public resources to more vulnerable subsectors, such as the forestry and logging sector. In contrast, the paper manufacturing industry presents a more pessimistic trajectory, with declines expected in employment, output, and the number of firms, while wages remain stagnant. Note that this trend is especially severe for Maine. These projections signal a need for proactive intervention to buffer communities at risk – through retraining programs, business development grants, or infrastructure investments that support industrial adaptation.

Given the regional diversity of the states analyzed – spanning the Northeast, North Central, South Central, and Pacific West – further research is warranted to explore regional economic differences in greater detail. Additionally, these findings have implications beyond the U.S., particularly for other resource-dependent economies such as Canada, Germany, Sweden, and Finland (Hetemaki et al., 2004; Malaty et al., 2006; Zhang & Parajuli, 2016). More broadly, the methodology used in this study can be adapted to other industries that serve as economic anchors for specific regions, such as mining, agriculture, and energy production (Andrei & Andrei, 2015; Apostu et al., 2022; Asnawi et al., 2025). This offers a valuable tool for analyzing economic resilience in industry-dependent communities and essential information for policy makers in these industries outside of the forestry sector. Evidence-based forecasting is especially vital in designing resilient policy responses tailored to local conditions at both the state and industry level.

## References

Al-Ballaa, N. R. (2005). Test for cointegration based on two-stage least squares. *Journal of Applied Statistics, 32*(7), 707–713. https://doi.org/10.1080/02664760500079571

Anderson, R. G., Hoffman, D. L., & Rasche, R. H. (2002). A vector error-correction forecasting model of the US economy. *Journal of Macroeconomics, 24*(4), 569-598.

Andrei, M., & Andrei, L. (2015). Vector Error Correction Model in Explaining the Association of Some Macroeconomic Variables in Romania. *Procedia Economics and Finance, 22,* 568-576. https://doi.org/10.1016/S2212-5671(15)00261-0

Apostu, S. A., Panait, M., Balsalobre-Lorente, D., Ferraz, D., & Rădulescu, I. G. (2022). Energy Transition in Non-Euro Countries from Central and Eastern Europe: Evidence from Panel Vector Error Correction Model. *Energies*, *15*(23), 9118. https://doi.org/10.3390/en15239118

Asnawi, Bachri, N. & Roni, M. (2025). Vector autoregressive model: to analyze the influence of the agriculture, mining and quarrying sectors on local revenues. *Miner Econ, 38,* 51–58. https://doi-org.ezproxy.lib.vt.edu/10.1007/s13563-024-00448-9

Baek, J. (2012). The long-run determinants of US lumber imported from Canada revisited. *Forest Policy and Economics, 14*, 69–73. https://doi.org/10.1016/j.forpol.2011.07.007.

Ballweg, J. (2017). Wisconsin Forestry Facts: Economic Impact. *Wisconsin Dept. Of Natural Resources*, https://dnr.wi.gov/about/documents/FactSheets/FactSheetForestryEconomy.pdf

Balogh, G. (2018, November 28). Timber Industry. *Encyclopedia of Arkansas*, https://encyclopediaofarkansas.net/entries/timber-industry-2143/

Banaś, J., & Utnik-Banaś, K. (2021). Evaluating a seasonal autoregressive moving average model with an exogenous variable for short-term timber price forecasting. *Forest Policy and Economics, 131*, 102564.

Bekhet, H. A., & Yusop, N. Y. M. (2009). Assessing the relationship between oil prices, energy consumption and macroeconomic performance in Malaysia: co-integration and vector error correction model (VECM) approach. *International business research, 2*(3), 152-175.

Bhatt, B. P., Chhetri, S. G., Silwal, T., & Poudel, M. (2021). Economic contribution of forestry sector to national economy in Nepal. *Journal of Resources and Ecology*, *12*(5), 620-627.

BLS. (2011). Using Location Quotients to Analyze Occupational Data. *BLS*, 1-13.

Bobeica, E., & Hartwig, B. (2023). The COVID-19 shock and challenges for inflation modelling. *International Journal of Forecasting, 39*(1), 519-539. https://doi.org/10.1016/j.ijforecast.2022.01.002

Cloughesy, M. (2021). Oregon Forest Facts & Figures. *Oregon Forest Resources Institute*. https://www.oregonloggers.org/docs/OR_Forest_Facts_and_Figures_2013.pdf

Conrad IV, J. L., Greene, W. D., & Hiesl, P. (2018). A review of changes in US logging businesses 1980s–present. *Journal of Forestry, 116*(3), 291-303.

Crandall, M., Anderson, J., & Rubin, J. (2017). Impacts of Recent Mill Closures and Potential Biofuels Development on Maine's Forest Products Industry. *Maine Policy Review, 26*(1), 15-22. https://digitalcommons.library.umaine.edu/mpr/vol26/iss1/4

Daigneault, A. J., Sohngen, B., & Kim, S. J. (2016). Estimating welfare effects from supply shocks with dynamic factor demand models. *Forest Policy and Economics, 73*, 41-51. https://doi.org/10.1016/j.forpol.2016.08.003

Eicher, T., Kuenzel, D., Papageorgiou, C., Christofides, C. (2019) Forecasts in times of crises. *International Journal of Forecasting, 35,* 1143-1159. https://doi.org/10.1016/j.ijforecast.2019.04.001

Engle, R. F., & Granger, W. J., (1987). Co-integration and error correction: representation, estimation and testing. *Econometrica, 55*(2), 251–276. https://www.jstor.org/stable/1913236

Federal Reserve System. (2016). The Federal Reserve System Purposes and Functions. *Federal Reserve System, 10*. https://www.federalreserve.gov/aboutthefed/files/pf_complete.pdf

Ferguene, M., Lehoux, N., & Dadouchi, C. (2023). Forecasting models for Quebec's lumber demand and exports using multivariate regression technique. *The Forestry Chronicle, 99*(1), 103-116.

Fickle, J. (2017, May). Forest Products Industry in Alabama. *Encyclopedia of Alabama*, http://encyclopediaofalabama.org/article/h-3021

Forest Opportunity Roadmap Maine. (2018, September). Vision and Roadmap for Maine's Forest Products Sector. https://formaine.org/

Gendron, J., Crawley, A., & Daigneault, A. (2019). Maine's Forestry and Logging Industry: Building a Model for Forecasting. University of Maine Electronic Theses and Dissertations. 2998. https://digitalcommons.library.umaine.edu/etd/2998

Guo, J., Presetemon, J., & Johnston, C. (2023). Forest market outlook in the Southern United States. *Forest Policy and Economics, 157*. https://doi.org/10.1016/j.forpol.2023.103091

Haji-Othman, M. S. (1991). Further assessment of the price competitiveness of Malaysian lauan lumber imports in the United States. *Forest Science, 37*(3), 849-859. https://doi.org/10.1093/forestscience/37.3.849

Henderson, J & Munn, I. (2008). Is the Forest Products Industry Important to Mississippi's Economy?. *Mississippi State University Extension*, http://extension.msstate.edu/content/the-forest-products-industry-important-mississippis-economy

Hernandez-Diaz, J., Musah, M., Morgan, T., Richey, R., Maggard, A., Via, B., Peresin, M. (2024). Wavelet Analysis and Forecasting using Open-Access Lumber Market Indices for Assessing the Impact of Hurricanes on Southern US Stumpage Prices. *Forest Products, 74*(1): 10-27. https://doi.org/10.13073/FPJ-D-23-00052

Hetemaki, L., Hanninen, R., & Toppinen, A. (2004). Short-Term Forecasting Models for the Finnish Forest Sector: Lumber Exports and Sawlog Demand. *Forest Science, 50*(4), 461-472. https://doi.org/10.1093/forestscience/50.4.461

Hietala, J., Hanninen, R. H., & Toppinen, A. (2013). Finnish and Swedish sawnwood exports to the UK market in the European monetary union regime. *Forest Science, 59*(4), 379–389. https://doi.org/10.5849/forsci.10-122

Hill, R. C., Griffiths, W. E., & Lim, G. C. (2008). Principles of econometrics. Hoboken, NJ: Wiley

Hodges, D. G., Hartsell, A. J., Brandeis, C., Brandeis, T. J., & Bentley, J. W. (2011). Recession effects on the forests and forest products industries of the South. *Forest products journal, 61*(8), 614-624.

Jiang, H., & Liu, C. (2014). A panel vector error correction approach to forecasting demand in regional construction markets. *Construction Management and Economics, 32*(12), 1205-1221.

Johansen, S., (1988). Statistical analysis of cointegration vectors. *Journal of Economic Dynamics and Control, 12*, 231–256. https://doi.org/10.1016/0165-1889(88)90041-3

Johansen, S., (1995). Likelihood-Based Inference in Cointegrated Vector Autoregressive Models. *Oxford*. https://doi-org.ezproxy.lib.vt.edu/10.1093/0198774508.001.0001

Kolo, H. & Tzanova, P. (2017). Forecasting the German forest products trade: A vector error correction model. *Journal of Forest Economics, 26*, 30-45. https://doi.org/10.1016/j.jfe.2016.11.001

Kozuch, A., Cywicka, D., & Adamowicz, K. (2023). A Comparison of Artificial Neural Network and Time Series Models for Timber Price Forecasting. *Forests 2023, 14*, 177.

Lambert, B. (2013). Error correction model. *Imperial College London.* Retrieved from https://www.youtube.com/watch?v=xVIkb-QeZ40

Lamichhane, S., Parajuli, R., Henderson, J. D., da Silva, B. K., & Tanger, S. M. (2025). Examining divergence in softwood lumber and timber prices in the southern United States. *Forest Science*, 1-20.

La, L. & Mei, B. (2015). Portfolio diversification through timber real estate investment trusts: A cointegration analysis. *Forest Policy and Economics, 1*, 347-355. https://doi.org/10.1016/j.forpol.2014.07.003

Liew, V. (2004). What lag selection criteria should we employ?, *Economics Bulletin, 33*(3), pp. 1-9.

Malaty, R., Toppinen, A., & Viitanen, J. (2006). Modelling and forecasting Finnish pine sawlog stumpage prices using alternative time-series methods. *Canadian Journal of Forest Research, 37*, 178-187. https://doi.org/10.1139/X06-208

Mei, B., Clutter, M., & Harris, T. (2010). Modeling and forecasting pine sawtimber stumpage prices in the US South by various time series models. *Canadian Journal of Forest Research, 40*, 1506-1516. https://doi.org/10.1139/X10-087

Munn, I. (1997). Logging - One of Mississippi's Most Important Industries. *Stephen F. Austin State University*, https://scholarworks.sfasu.edu/cgi/viewcontent.cgi?article=1164&context=forestry

Michieka, N. M., & Gearhart, R. (2015). Oil price fluctuations and employment in Kern County: A Vector Error Correction approach. *Energy Policy, 87*, 584-590.

Nagubadi, R. V., & Zhang, D. (2013). US imports for Canadian softwood lumber in the context of trade dispute: A cointegration approach. *Forest Science, 59*(5), 517–523. https://doi.org/10.5849/forsci.12-013

Nanang, D. M. (2000). A multivariate cointegration test of the law of one price for Canadian softwood lumber markets. *Forest Policy and Economics, 1*, 347-355. https://doi.org/10.1016/S1389-9341(00)00028-9

Nanang, D. M. (2009). Analysis of export demand for Ghana's timber products: A multivariate co-integration approach. *Journal of Forest Economics, 16*, 47-61. https://dx.doi.org/10.1016/j.jfe.2009.06.001

Parajuli, R., & Chang, S. J. (2015). The softwood sawtimber stumpage market in Louisiana: market dynamics, structural break, and vector error correction model. *Forest Science, 61*(5). 904-913. https://doi.org/10.5849/forsci.14-099

Parajuli, R., Zhang, D., & Chang, S. J. (2016). Modeling stumpage markets using vector error correction vs. simultaneous equation estimation approach: A case of the Louisiana sawtimber market. *Forest Policy and Economics, 70*, 16-19. https://doi.org/10.1016/j.forpol.2016.05.013

Pattanayak, S. K., Murray, B. C., & Abt, R. C. (2002). How Joint is Joint Forest Production? An Econometric Analysis of Timber Supply Conditional on Endogenous Amenity Values. *Forest Science, 48*(3), 479-491. https://doi.org/10.1093/forestscience/48.3.479

Pelkki, M. (2005). An Economic Assessment of Arkansas' Forest Industries. *University of Arkansas-Monticello*, https://agcomm.uark.edu/agnews/publications/afrc-007.2.1-20.pdf

Polemis, M. (2007). Modeling industrial energy demand in Greece using cointegration techniques. *Energy Policy, 35*, 4039-4050. https://doi.org/10.1016/j.enpol.2007.02.007

Poudel, J., & Dahal, R. (2025). A comprehensive look at the forest products industry's economic contribution to the United States: Pre-and post-COVID analysis. *Forest Policy and Economics, 172*, 103440.

Robbins, W. (2021, Jan 20). Timber Industry. O*regon Encyclopedia*, https://www.oregonencyclopedia.org/articles/timber_industry/#.YHnRZmdKhEY

Schorfheide, F. & Song, D. (2024). Real-Time Forecasting with a (Standard) Mixed-Frequency VAR During a Pandemic. *International Journal of Central Banking, 20*(4), 275-320. https://doi.org/10.3386/w29535

Sherman, G., & Daigneault, A. (2023). Evaluation of Maine Resident Perceptions on Community Resilience, Conservation, and Natural Resource Industries. *Society & Natural Resources, 36*(3), 211-231.

Song, N., Chang, S. J., & Aguilar, F. X. (2011). U.S. softwood lumber demand and supply estimation using cointegration in dynamic equations. *Journal of Forest Economics, 17*, 19-33. https://dx.doi.org/10.1016/j.jfe.2010.07.002

Song, N., Aguilar, F. X., Shifley, S. R., & Goerndt, M. E. (2012). Analysis of U.S. residential wood energy consumption: 1967-2009. *Energy Economics, 34*, 2116-2124. https://doi.org/10.1016/j.eneco.2012.03.004

Stordal, S., & Nyrud, A. Q. (2003). Testing roundwood market efficiency using a multivariate cointegration estimator. *Forest Policy and Economics, 5*, 57-68. https://doi.org/10.1016/S1389-9341(02)00015-1

Tanger, S & Measells, M. (2020). The Economic Contributions of Forestry and Forest Products - Mississippi. *Mississippi State University Extension*, http://extension.msstate.edu/publications/the-economic-contributions-forestry-and-forest-products-mississippi

Toppinen, A. (1998). Incorporating cointegration relations in a short-run model of the Finnish sawlog market. *Canadian Journal of Forest Research, 28*, 29

Toppinen, A., & Kuuluvainen, J. (2010). Forest sector modelling in Europe—the state of the art and future research directions. *Forest Policy and Economics, 12*, 2-8. http://doi.org/10.1016/j.forpol.2009.09.017

U.S. Department of Agriculture. (2011). National report on sustainable forests, 2010. *USDA, Forest Service*.

Woodall, C. W., Ince, P. J., Skog, K. E., Aguilar, F. X., Keegan, C. E., Sorenson, C. B., ... & Smith, W. B. (2011). An overview of the forest products sector downturn in the United States. *Forest Products Journal, 61*(8), 595-603.

Xu, J., Tao, R., & Amacher, G. S. (2004). An empirical analysis of China's state-owned forests. *Forest Policy and Economics, 6*, 379-390. https://doi.org/10.1016/j.forpol.2004.03.013

Yin, R., & Baek, J. (2005). Is there a single national lumber market in the United States? *Forest Science, 51*(2), 155–164. https://doi.org/10.1093/forestscience/51.2.155

Yin, R., & Xu, J. (2003). Identifying the inter-market relationships of forest products in the Pacific Northwest with cointegration and causality tests. *Forest Policy and Economics, 5*, 305-316.

Zhang, D., & Parajuli, R. (2016). Policy impacts estimates are sensitive to data selection in empirical analysis: evidence from the United States – Canada softwood lumber trade dispute. *Canadian Journal of Forest Research, 46*, 1343-1347. https://doi.org/10.1139/cjfr-2016-0168

Zhang, Y., & Xie, H. (2019). Interactive Relationship among Urban Expansion, Economic Development, and Population Growth since the Reform and Opening up in China: An Analysis Based on a Vector Error Correction Model. *Land, 8*, 153. https://doi.org/10.3390/land8100153

**Appendix A.** Model summary statistics and robustness checks

**Table A1.** Summary statistics of key variables used to model each state for NAICS 113.

| State | Variable | N | Mean | Standard Deviation | Minimum | Maximum |
|---|---|---|---|---|---|---|
| **N/A** | Price | 72 | 0.888 | 0.070 | 0.762 | 1.010 |
| | Quarter | 72 | n/a | n/a | 2001Q1 | 2018Q4 |
| **Alabama** | Employment (thousands) | 72 | 4770.722 | 528.868 | 4059 | 5852 |
| | Output (millions) | 72 | 138.364 | 10.343 | 125.925 | 167.450 |
| | Wages | 72 | 40045.820 | 3669.998 | 32128 | 48561 |
| | Number of Firms | 72 | 723.500 | 102.239 | 618 | 958 |
| **Arkansas** | Employment (thousands) | 72 | 2634.806 | 486.663 | 2080 | 3490 |
| | Output (millions) | 72 | 139.816 | 8.399 | 127.600 | 158.175 |
| | Wages | 72 | 21729.430 | 2461.600 | 16572 | 27078 |
| | Number of Firms | 72 | 488.889 | 87.577 | 395 | 637 |
| **Maine** | Employment (thousands) | 72 | 2435.694 | 402.883 | 1532 | 2945 |
| | Output (millions) | 72 | 126.094 | 25.333 | 95.250 | 184.046 |
| | Wages | 72 | 22862.690 | 4219.346 | 13490 | 28600 |
| | Number of Firms | 72 | 465.167 | 32.614 | 410 | 539 |
| **Mississippi** | Employment (thousands) | 72 | 3509.125 | 525.034 | 2916 | 4367 |
| | Output (millions) | 72 | 131.736 | 15.187 | 109.543 | 163.417 |
| | Wages | 72 | 27146.690 | 2290.516 | 22072 | 31972 |
| | Number of Firms | 72 | 543.347 | 83.891 | 445 | 696 |
| **Oregon** | Employment (thousands) | 72 | 6520.056 | 914.163 | 5024 | 7931 |
| | Output (millions) | 72 | 371.786 | 35.719 | 316.446 | 443.350 |
| | Wages | 72 | 74452.690 | 9361.837 | 48541 | 95042 |
| | Number of Firms | 72 | 779.542 | 116.811 | 653 | 1009 |

**Table A2.** Summary statistics of key variables used to model each state for NAICS 321.

| State | Variable | N | Mean | Standard Deviation | Minimum | Maximum |
|---|---|---|---|---|---|---|
| **N/A** | Price | 60 | 0.844 | 0.080 | 0.700 | 1.042 |
| | Quarter | 60 | n/a | n/a | 2004Q1 | 2018Q4 |
| **Alabama** | Employment (thousands) | 60 | 16290.200 | 3211.876 | 12124 | 21922 |
| | Output (millions) | 60 | 249.012 | 36.851 | 164.080 | 318.712 |
| | Wages | 60 | 150755.500 | 28499.020 | 95805 | 208886 |
| | Number of Firms | 60 | 408.417 | 48.883 | 351 | 483 |
| **Arkansas** | Employment (thousands) | 60 | 10505.350 | 1901.429 | 8607 | 14003 |
| | Output (millions) | 60 | 179.579 | 25.049 | 131.518 | 221.631 |
| | Wages | 60 | 93793.170 | 11820.660 | 72067 | 113987 |
| | Number of Firms | 60 | 348.233 | 33.143 | 300 | 403 |
| **Maine** | Employment (thousands) | 60 | 4915.300 | 941.203 | 3902 | 6840 |
| | Output (millions) | 60 | 78.765 | 9.636 | 58.651 | 102.044 |
| | Wages | 60 | 45630.900 | 6859.528 | 31724 | 59522 |
| | Number of Firms | 60 | 217.167 | 31.553 | 181 | 283 |
| **Mississippi** | Employment (thousands) | 60 | 10600.500 | 2199.874 | 8512 | 14465 |
| | Output (millions) | 60 | 183.422 | 24.604 | 129.520 | 228.850 |
| | Wages | 60 | 97188.620 | 14736.430 | 70435 | 127558 |
| | Number of Firms | 60 | 283.767 | 27.526 | 244 | 325 |
| **Oregon** | Employment (thousands) | 60 | 24457.850 | 4868.420 | 18917 | 32858 |
| | Output (millions) | 60 | 427.237 | 50.814 | 302.302 | 504.151 |
| | Wages | 60 | 261847.300 | 41054.910 | 190597 | 327597 |
| | Number of Firms | 60 | 462.117 | 44.934 | 407 | 533 |
| **Wisconsin** | Employment (thousands) | 60 | 19346.480 | 3852.204 | 15661 | 26103 |
| | Output (millions) | 60 | 272.365 | 25.497 | 226.400 | 319.415 |
| | Wages | 60 | 165318.400 | 27961.030 | 109319 | 216752 |
| | Number of Firms | 60 | 594.450 | 61.748 | 518 | 686 |

**Table A3.** Summary statistics of key variables used to model each state for NAICS 322.

| State | Variable | N | Mean | Standard Deviation | Minimum | Maximum |
|---|---|---|---|---|---|---|
| **N/A** | Price | 60 | 0.900 | 0.086 | 0.700 | 1.034 |
| | Quarter | 60 | n/a | n/a | 2004Q1 | 2018Q4 |
| **Alabama** | Employment (thousands) | 60 | 12719.620 | 1261.083 | 11025 | 14906 |
| | Output (millions) | 60 | 709.239 | 90.370 | 570.600 | 890.713 |
| | Wages | 60 | 229630.800 | 13505.820 | 204766 | 277691 |
| | Number of Firms | 60 | 98.917 | 7.552 | 88 | 116 |
| **Arkansas** | Employment (thousands) | 60 | 10567.650 | 946.624 | 9311 | 12408 |
| | Output (millions) | 60 | 456.539 | 51.549 | 375.600 | 561.713 |
| | Wages | 60 | 151272.800 | 7235.220 | 136843 | 167105 |
| | Number of Firms | 60 | 79.533 | 6.342 | 69 | 92 |
| **Maine** | Employment (thousands) | 60 | 7159.017 | 1752.367 | 4303 | 10141 |
| | Output (millions) | 60 | 251.828 | 79.617 | 147.475 | 391.131 |
| | Wages | 60 | 114647.900 | 18452.910 | 80014 | 145207 |
| | Number of Firms | 60 | 29.300 | 7.911 | 19 | 47 |
| **Mississippi** | Employment (thousands) | 60 | 4392.867 | 723.909 | 3595 | 5840 |
| | Output (millions) | 60 | 172.613 | 28.778 | 133.050 | 241.616 |
| | Wages | 60 | 64160.980 | 7299.291 | 53038 | 82394 |
| | Number of Firms | 60 | 62.983 | 4.545 | 58 | 74 |
| **Wisconsin** | Employment (thousands) | 60 | 32667.700 | 3009.017 | 29441 | 38962 |
| | Output (millions) | 60 | 1172.130 | 199.072 | 912.375 | 1553.405 |
| | Wages | 60 | 481891.300 | 22935.080 | 427142 | 521102 |
| | Number of Firms | 60 | 268.467 | 19.314 | 244 | 302 |

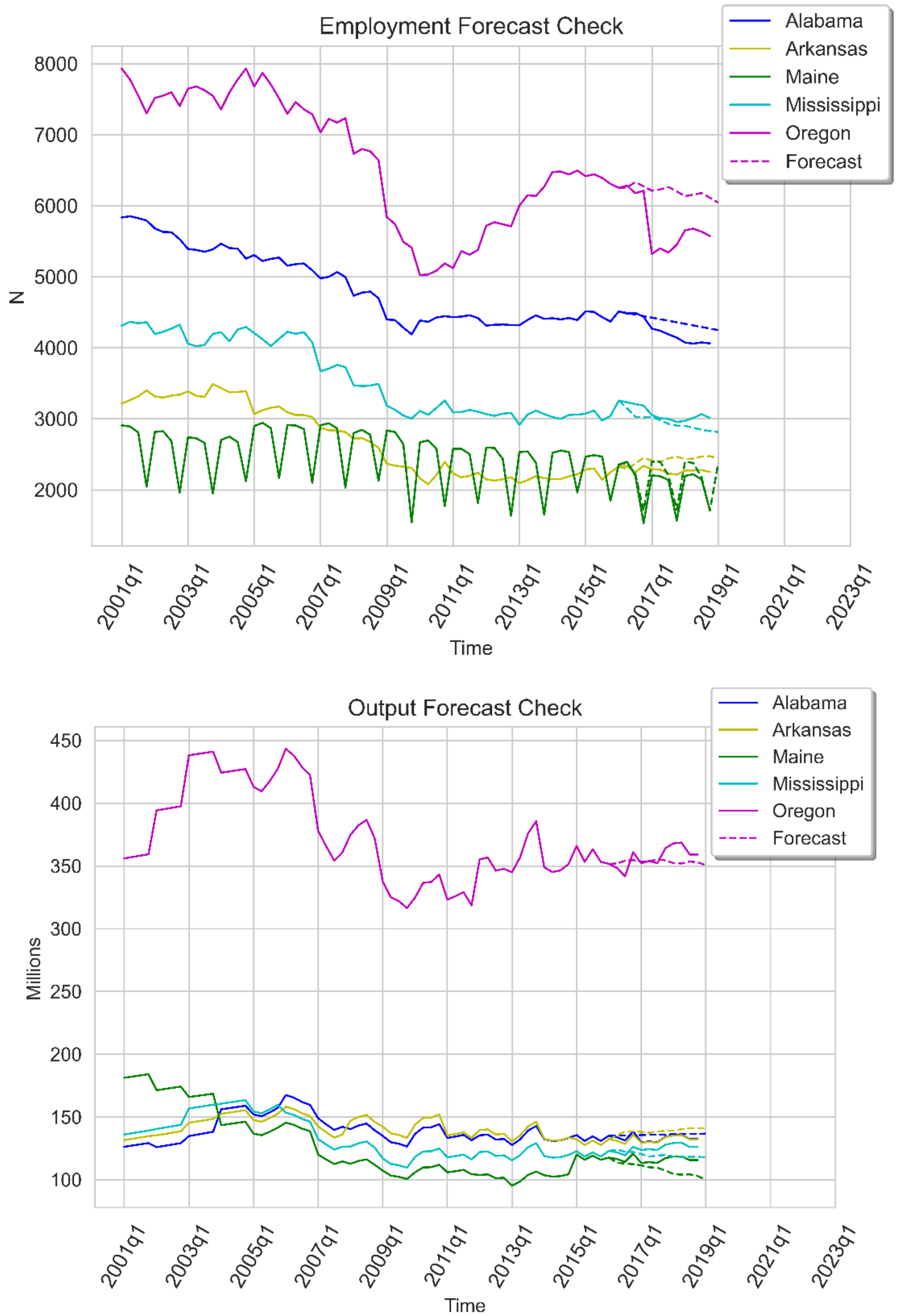

**Figure A1.** NAICS 113 robustness check for employment and output comparing actual to forecasted values, 2016 - 2018.

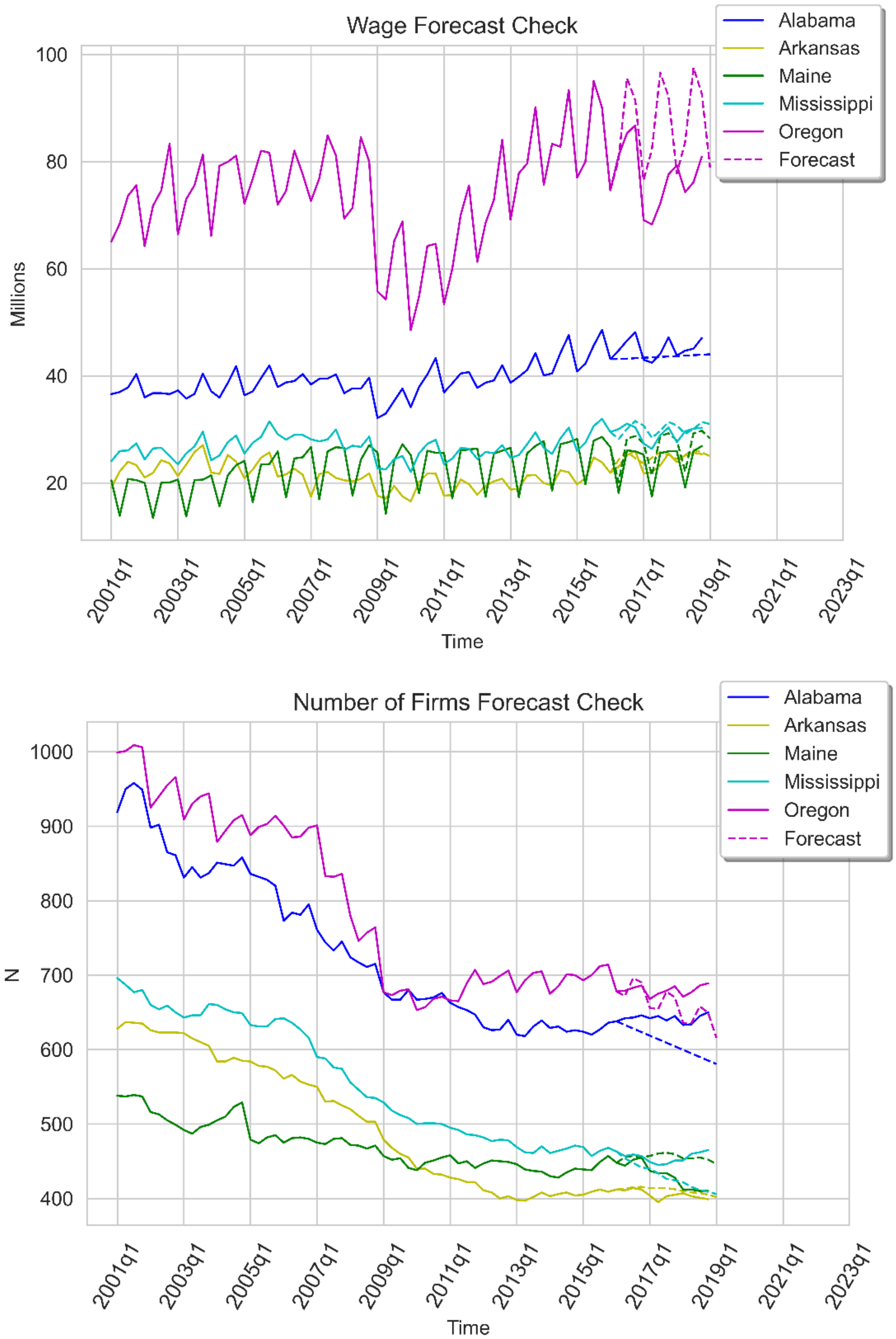

**Figure A2.** NAICS 113 robustness check for wage and number of firms comparing actual to forecasted values, 2016 - 2018.

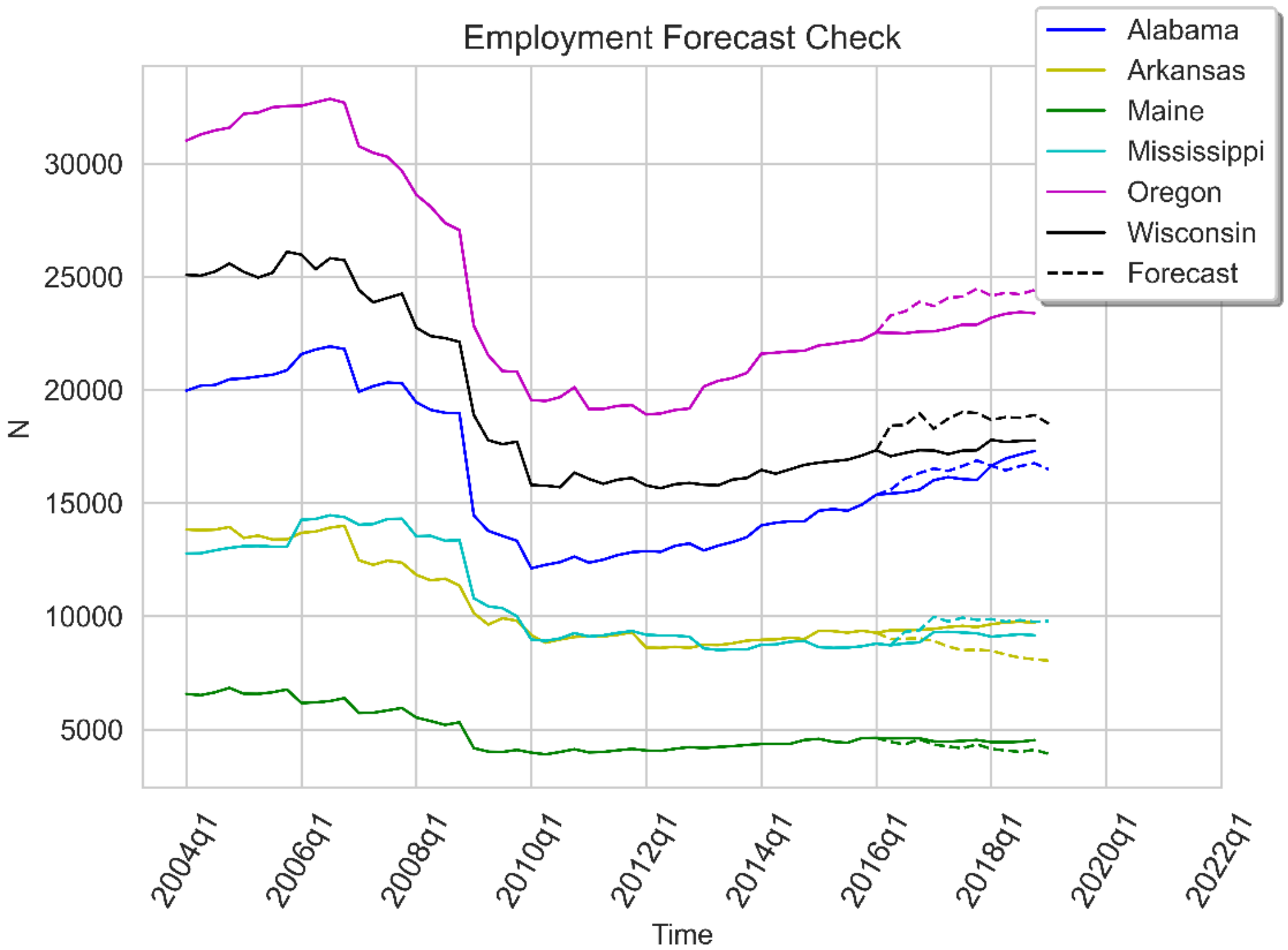

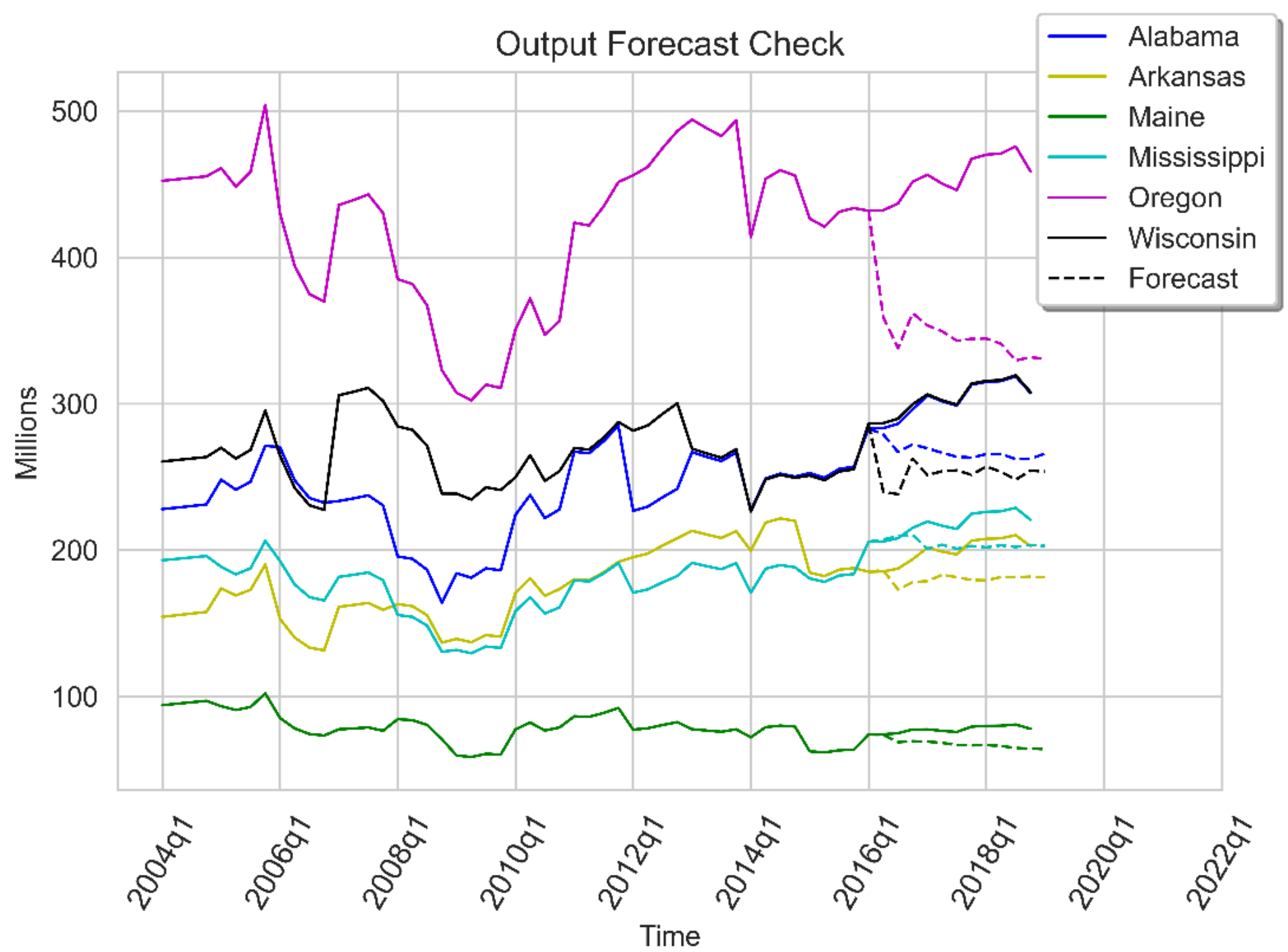

**Figure A3.** NAICS 321 robustness check for employment and output comparing actual to forecasted values, 2016 - 2018.

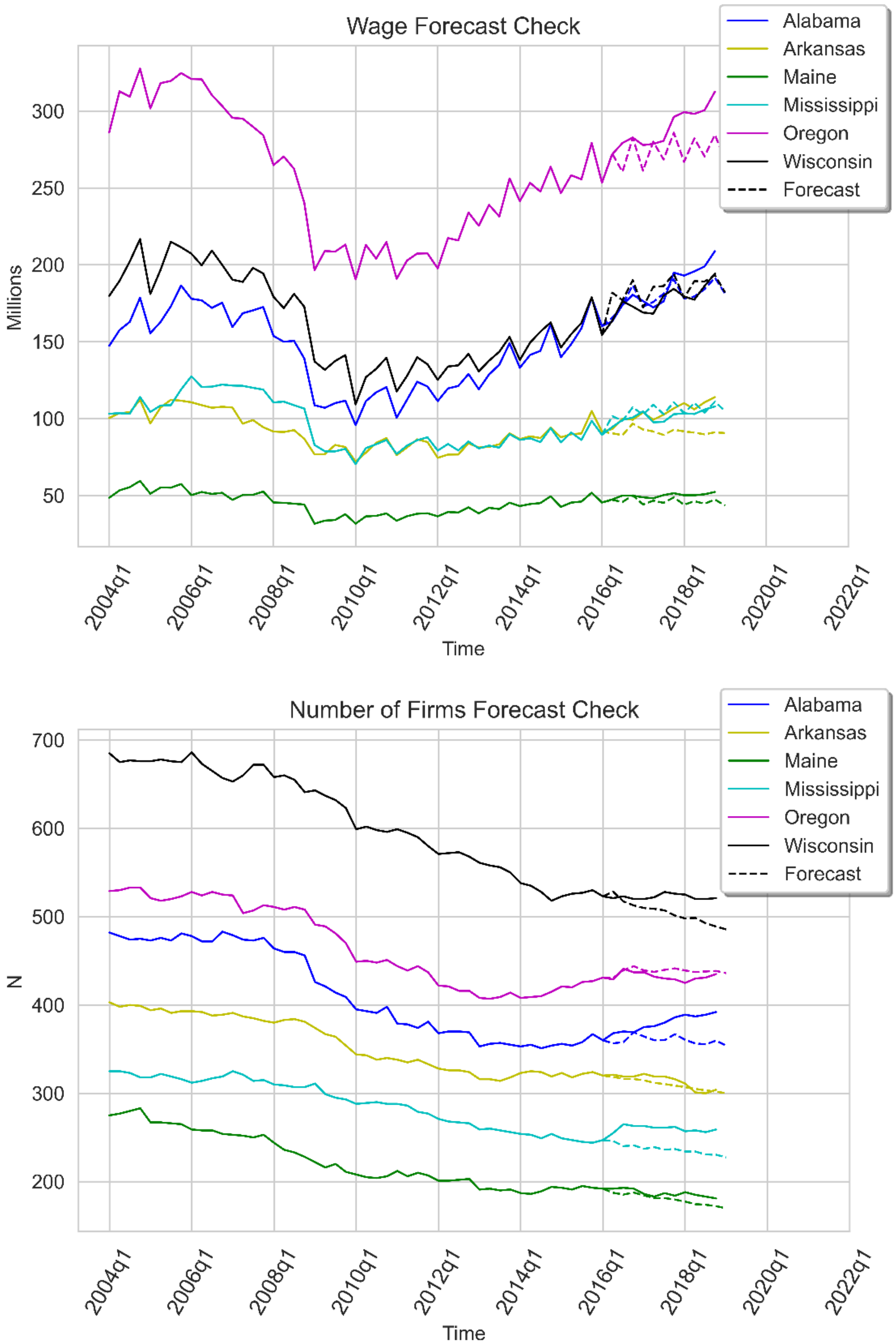

**Figure A4.** NAICS 321 robustness check for wage and number of firms comparing actual to forecasted values, 2016 - 2018.

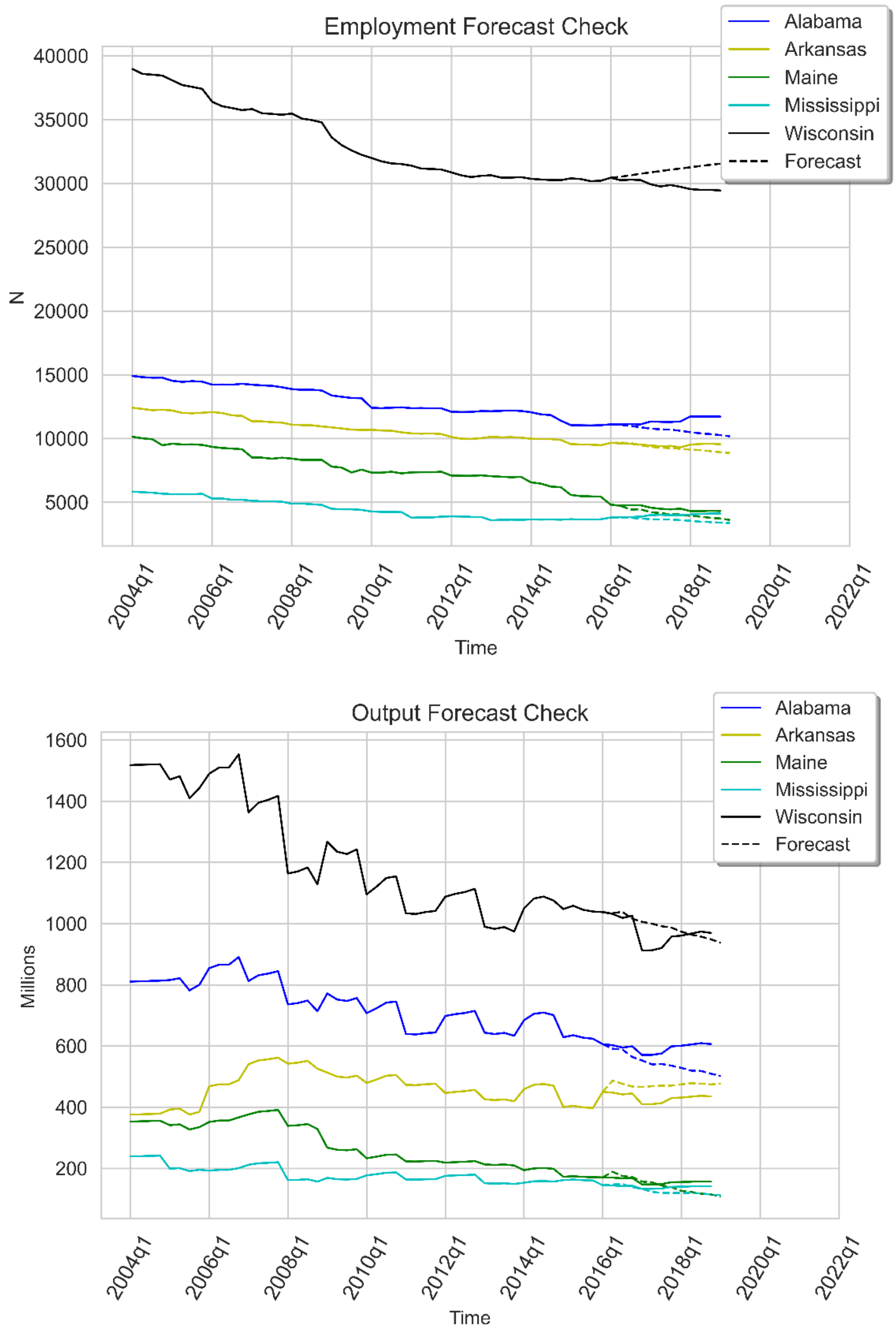

**Figure A5.** NAICS 322 robustness check for employment and output comparing actual to forecasted values, 2016 - 2018.

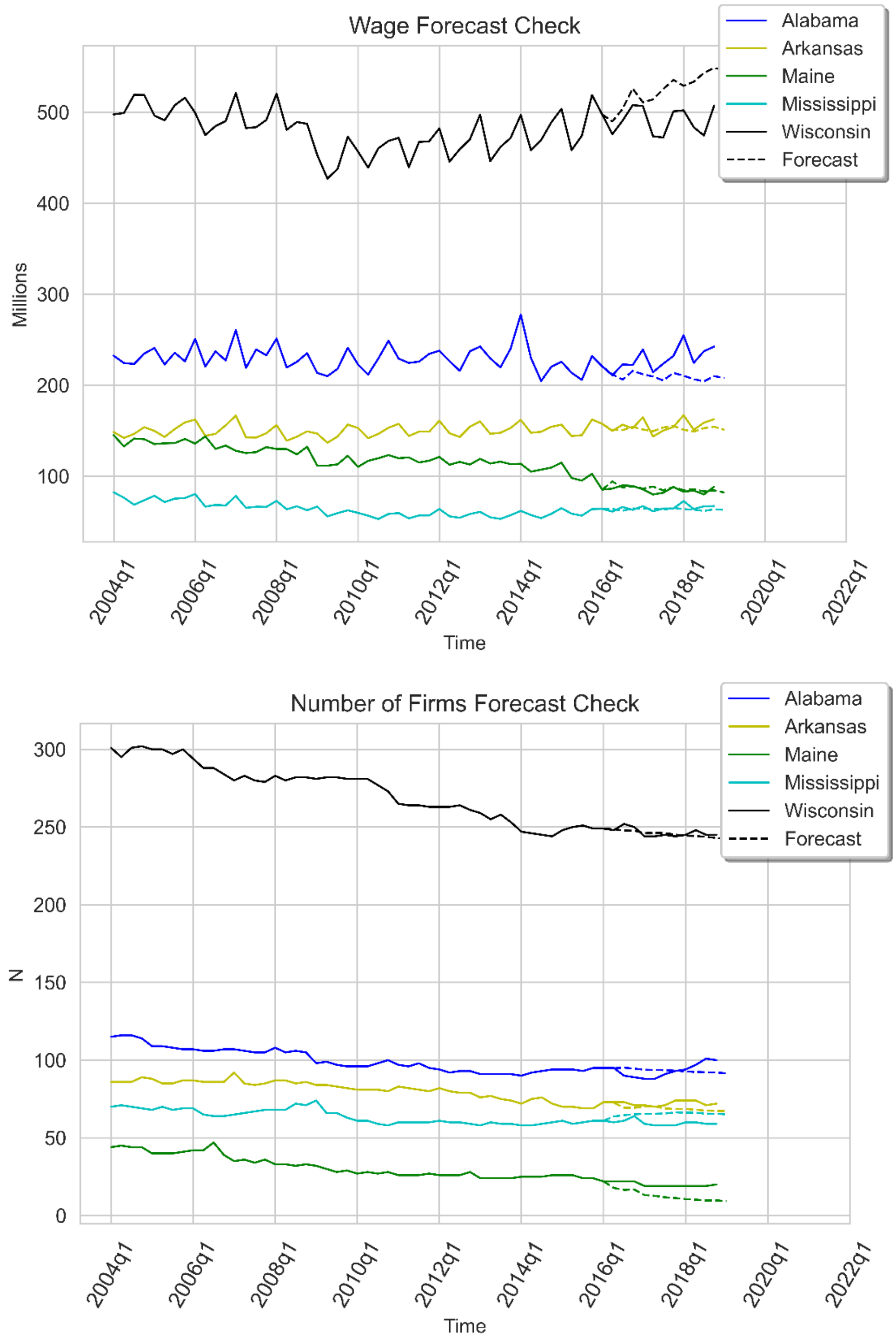

**Figure A6.** NAICS 322 robustness check for wage and number of firms comparing actual to forecasted values, 2016 - 2018.

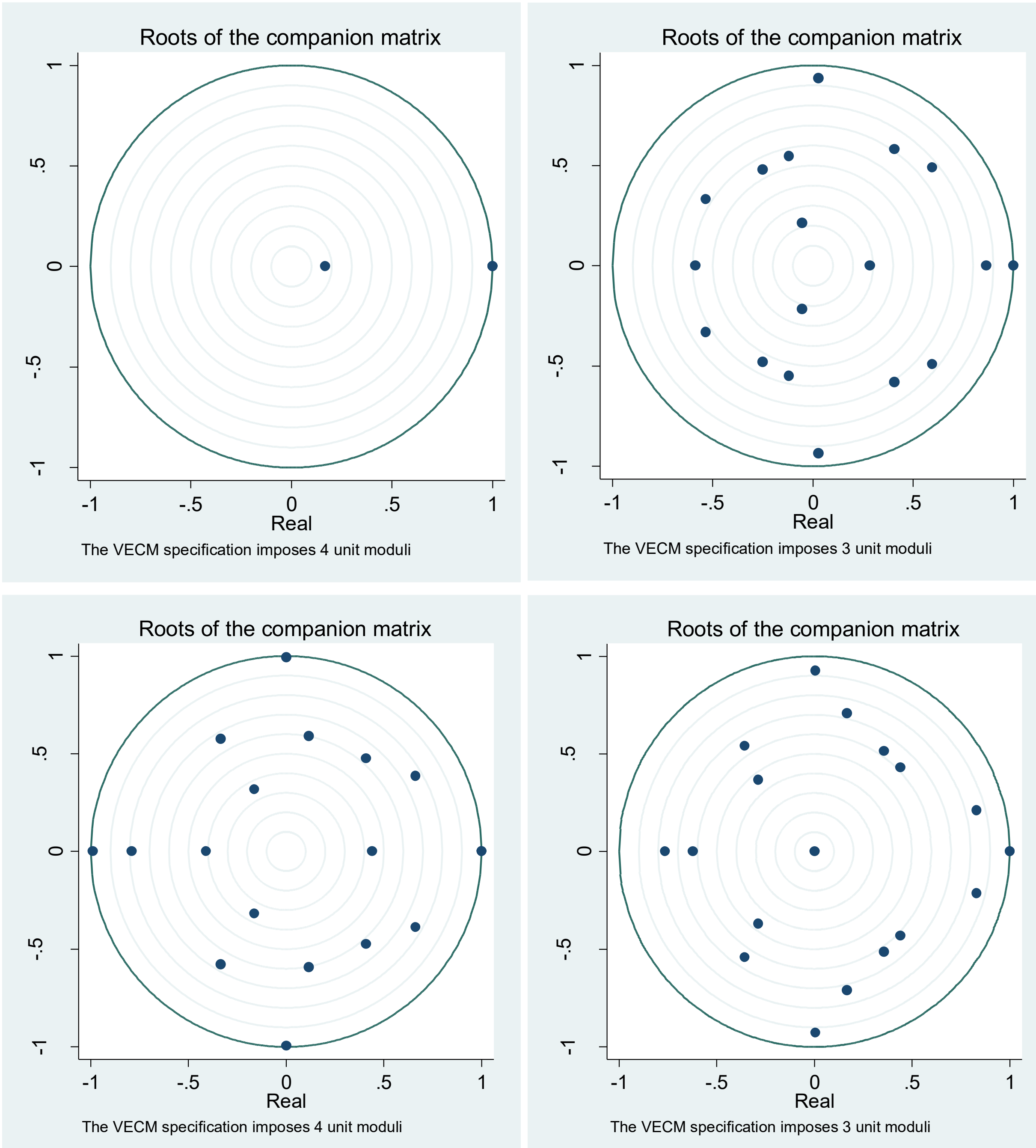

**Figure A7.** NAICS 113 trend stability robustness check. Top left to bottom right: AL, AR, ME, MS.

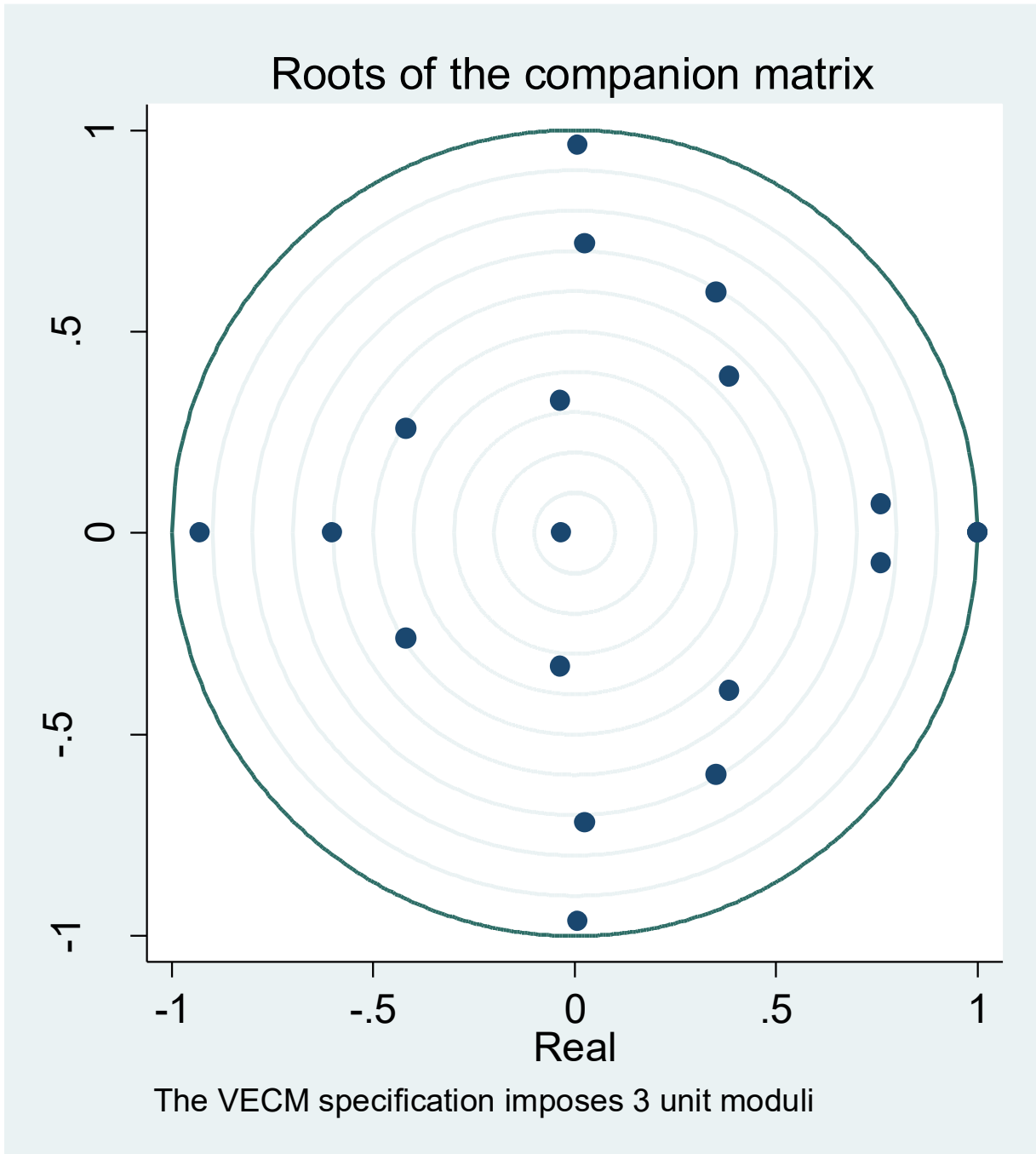

**Figure A8.** NAICS 113 trend stability robustness check. OR.

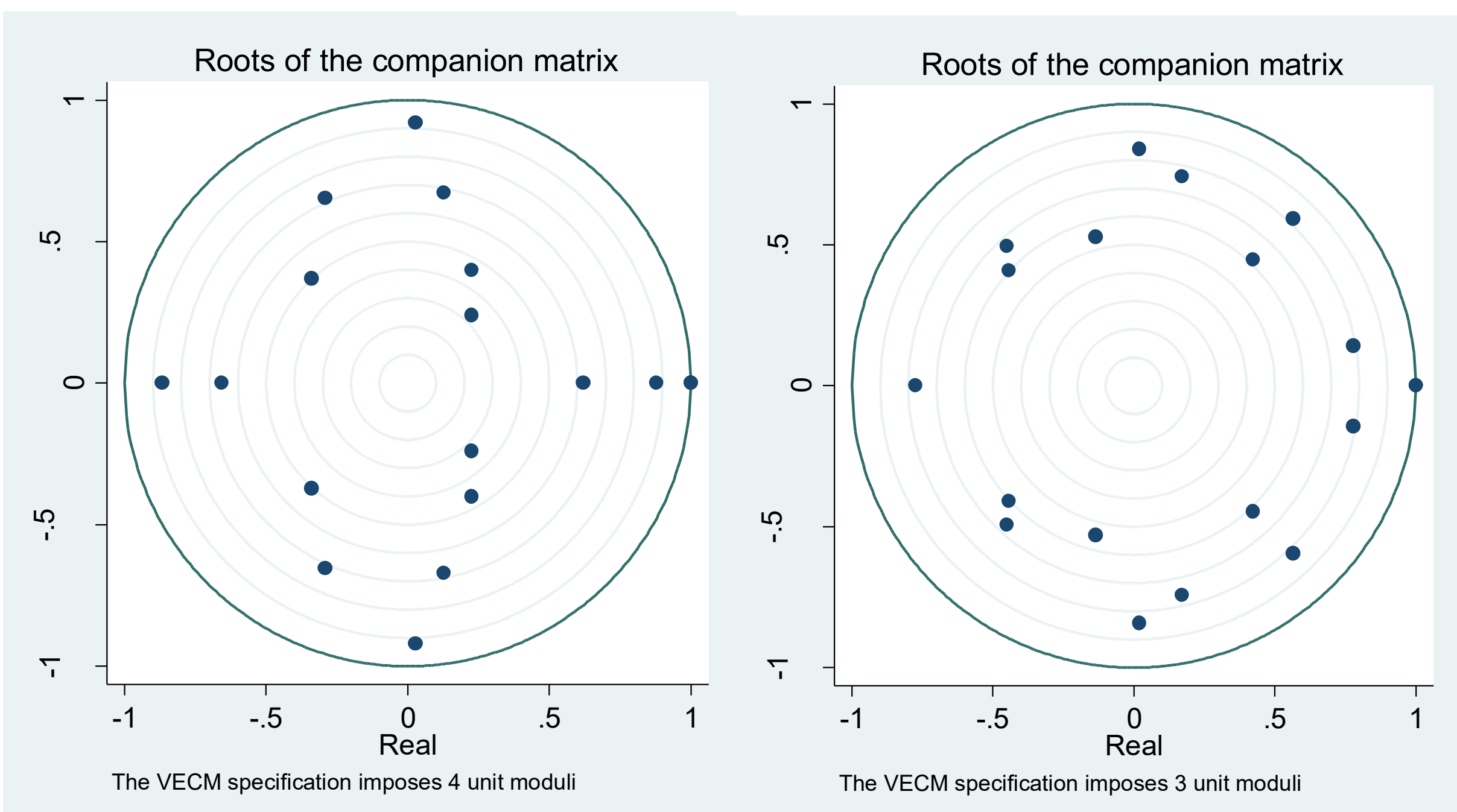

**Figure A9.** NAICS 321 trend stability robustness check. Left to right: AL, AR.

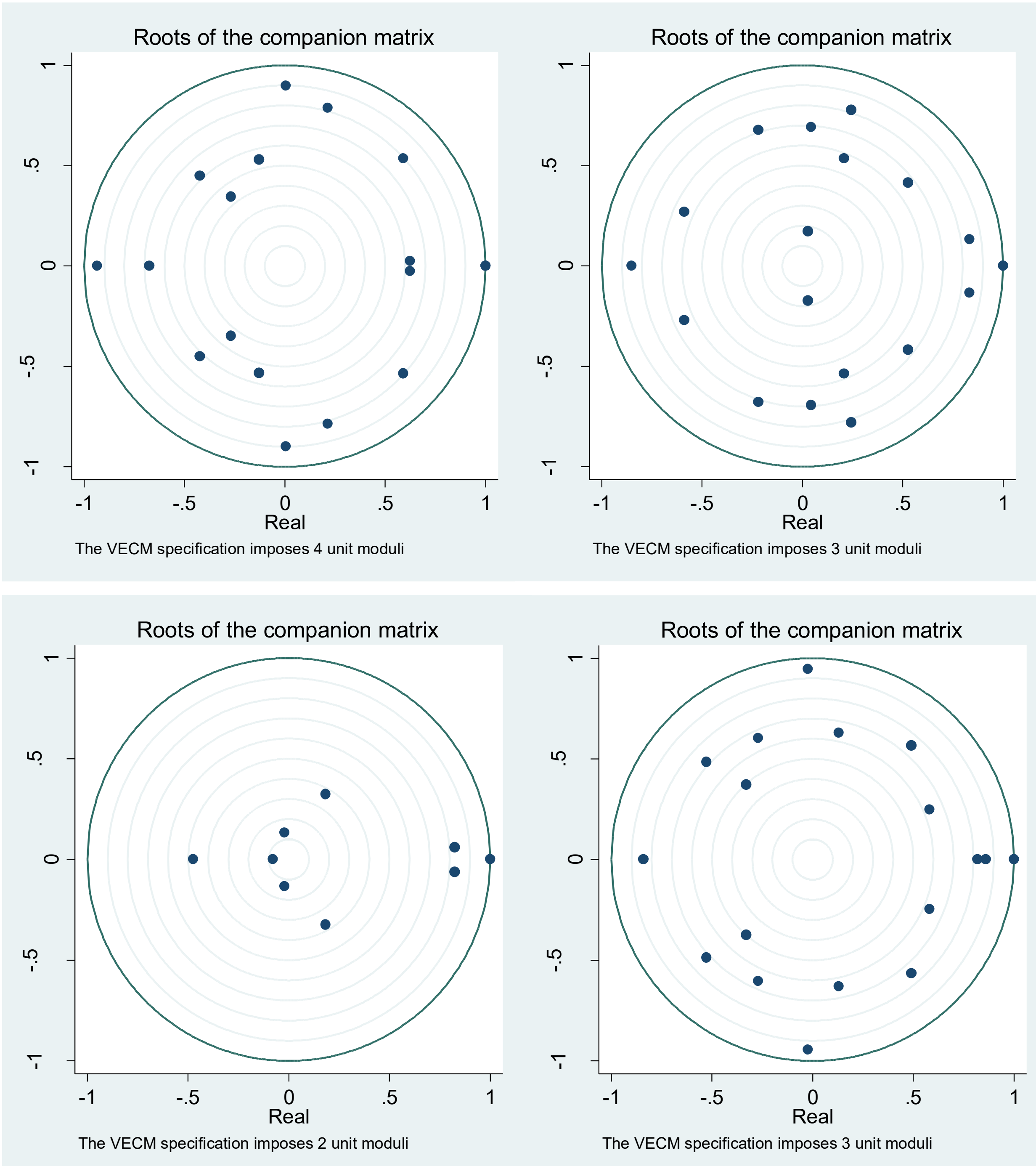

**Figure A10.** NAICS 321 trend stability robustness check. Top left to bottom right: ME, MS, OR, WI.

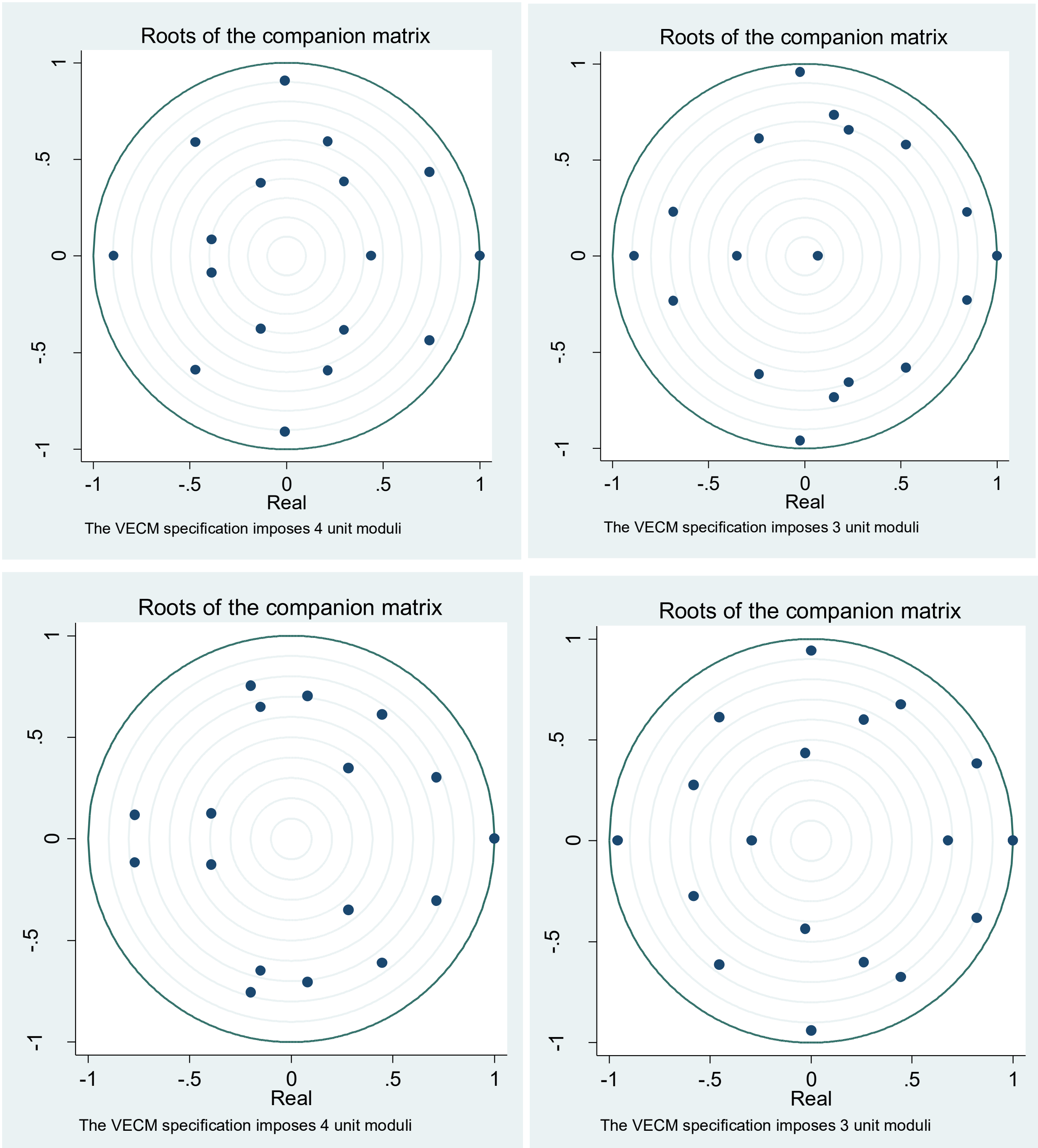

**Figure A11.** NAICS 322 trend stability robustness check. Top left to bottom right: AL, AR, ME, MS.

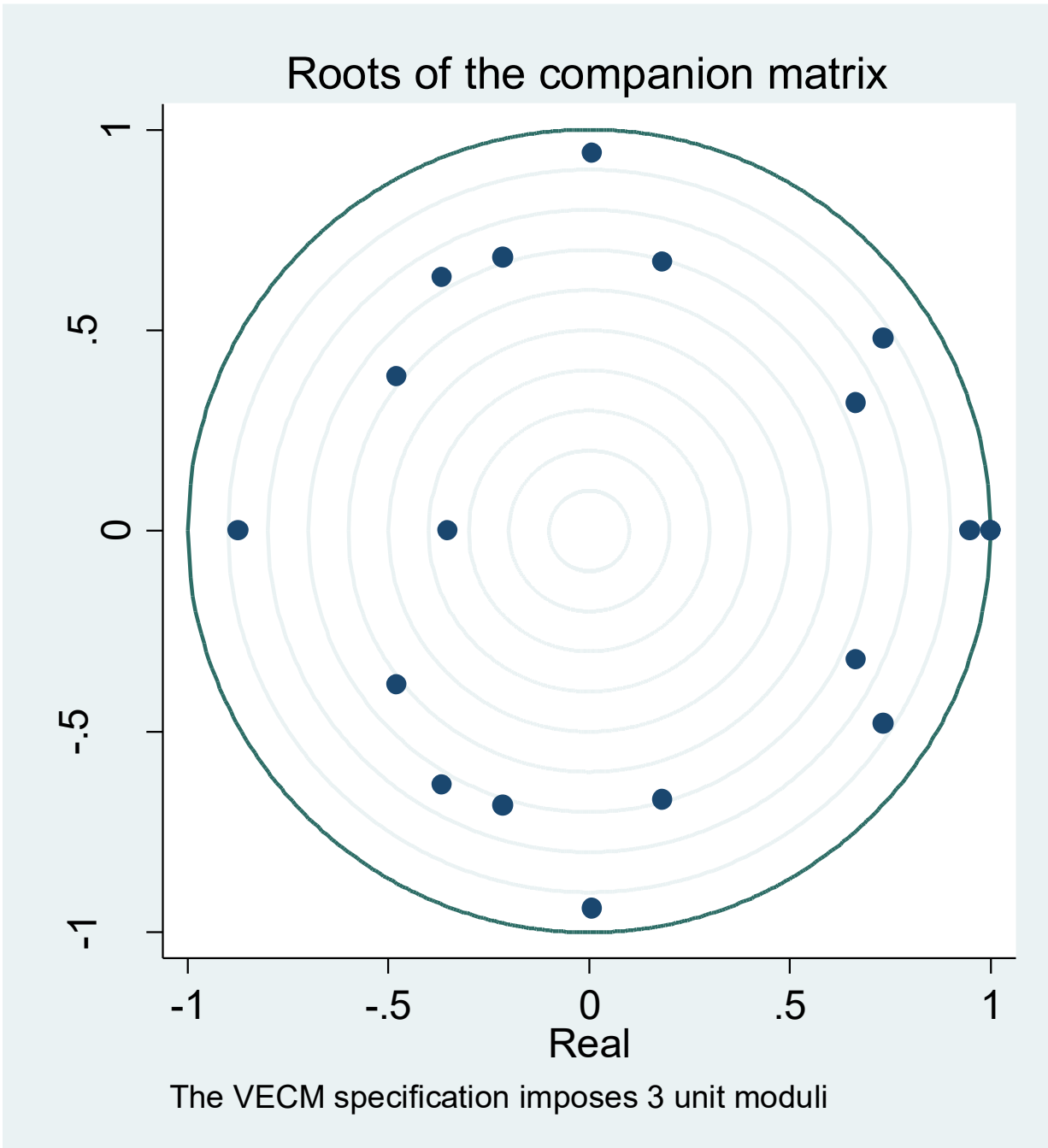

**Figure A12.** NAICS 322 trend stability robustness check. WI.

**Appendix B.** Model testing: Lags, Cointegration, Autocorrelation and Normality

**Table B1.** Summary of Model Tests. Includes rank, lag length, and normality for each NAICS.

| NAICS | State | Cointegration Rank | Lag Length | Normally Distributed? |
|---|---|---|---|---|
| 113 | AL | 1 | 1 | Y |
| | AR | 2 | 4 | N |
| | ME | 1 | 4 | Y |
| | MS | 2 | 4 | N |
| | OR | 2 | 4 | Y |
| 321 | AL | 1 | 4 | Y |
| | AR | 2 | 4 | Y |
| | ME | 1 | 4 | N |
| | MS | 2 | 4 | Y |
| | OR | 2 | 3 | Y |
| | WI | 2 | 4 | Y |
| 322 | AL | 1 | 4 | Y |
| | AR | 2 | 4 | Y |
| | ME | 1 | 4 | Y |
| | MS | 1 | 4 | Y |
| | WI | 2 | 4 | N |

**Table B2.** Normality Tests. NAICS 113. Performed for our VEC model, these include Jarque-Bera, Skewness, and Kurtosis.

| Normality Test (Alabama) | | | | | | | | | | | |
|---|---|---|---|---|---|---|---|---|---|---|---|
| Test Type | Jarque-Bera Test | | | Skewness | | | | Kurtosis | | | |
| Equation | $\chi^2$ | Df | P-Value | Skew | $\chi^2$ | Df | P-Value | Kurtosis | $\chi^2$ | Df | P-Value |
| D_output | 3.987 | 2 | 0.136 | 0.263 | 0.795 | 1 | 0.373 | 4.054 | 3.192 | 1 | 0.074 |
| D_price | 16.960 | 2 | 0.000 | 0.435 | 2.181 | 1 | 0.140 | 5.267 | 14.779 | 1 | 0.000 |
| D_employment | 2.579 | 2 | 0.275 | -0.110 | 0.139 | 1 | 0.709 | 3.921 | 2.440 | 1 | 0.118 |
| D_wages | 0.141 | 2 | 0.931 | 0.049 | 0.028 | 1 | 0.868 | 2.801 | 0.114 | 1 | 0.736 |
| D_numFirms | 13.726 | 2 | 0.001 | -0.782 | 7.033 | 1 | 0.008 | 4.526 | 6.693 | 1 | 0.010 |
| ALL | 37.39 | 10 | 0.000 | | 10.18 | 5 | 0.070 | | 27.22 | 5 | 0.000 |

| Normality Test (Arkansas) | | | | | | | | | | | |
|---|---|---|---|---|---|---|---|---|---|---|---|
| Test Type | Jarque-Bera Test | | | Skewness | | | | Kurtosis | | | |
| Equation | $\chi^2$ | Df | P-Value | Skew | $\chi^2$ | Df | P-Value | Kurtosis | $\chi^2$ | Df | P-Value |
| D_output | 1.723 | 2 | 0.423 | -0.340 | 1.312 | 1 | 0.252 | 3.381 | 0.411 | 1 | 0.522 |
| D_price | 8.917 | 2 | 0.012 | 0.301 | 1.029 | 1 | 0.310 | 4.669 | 7.888 | 1 | 0.005 |
| D_employment | 0.720 | 2 | 0.698 | -0.117 | 0.156 | 1 | 0.693 | 2.554 | 0.564 | 1 | 0.453 |
| D_wages | 1.288 | 2 | 0.525 | 0.164 | 0.304 | 1 | 0.582 | 2.411 | 0.984 | 1 | 0.321 |
| D_numFirms | 5.696 | 2 | 0.058 | -0.457 | 2.363 | 1 | 0.124 | 4.085 | 3.333 | 1 | 0.068 |
| ALL | 18.34 | 10 | 0.049 | | 5.16 | 5 | 0.396 | | 13.18 | 5 | 0.022 |

| Normality Test (Maine) | | | | | | | | | | | |
|---|---|---|---|---|---|---|---|---|---|---|---|
| Test Type | Jarque-Bera Test | | | Skewness | | | | Kurtosis | | | |
| Equation | $\chi^2$ | Df | P-Value | Skew | $\chi^2$ | Df | P-Value | Kurtosis | $\chi^2$ | Df | P-Value |
| D_output | 35.175 | 2 | 0.000 | -0.845 | 6.776 | 1 | 0.009 | 6.458 | 28.399 | 1 | 0.000 |
| D_price | 2.392 | 2 | 0.302 | -0.118 | 0.133 | 1 | 0.716 | 3.975 | 2.260 | 1 | 0.133 |
| D_employment | 59.672 | 2 | 0.000 | 0.874 | 7.249 | 1 | 0.007 | 7.698 | 52.423 | 1 | 0.000 |
| D_wages | 1.476 | 2 | 0.478 | -0.391 | 1.450 | 1 | 0.228 | 2.895 | 0.026 | 1 | 0.872 |
| D_numFirms | 25.242 | 2 | 0.000 | -0.714 | 4.841 | 1 | 0.028 | 5.931 | 20.401 | 1 | 0.000 |
| ALL | 121.96 | 10 | 0.000 | | 20.45 | 5 | 0.001 | | 103.51 | 5 | 0.000 |

| Normality Test (Mississippi) | | | | | | | | | | | |
|---|---|---|---|---|---|---|---|---|---|---|---|
| Test Type | Jarque-Bera Test | | | Skewness | | | | Kurtosis | | | |
| Equation | $\chi^2$ | Df | P-Value | Skew | $\chi^2$ | Df | P-Value | Kurtosis | $\chi^2$ | Df | P-Value |
| D_output | 0.621 | 2 | 0.733 | -0.136 | 0.208 | 1 | 0.648 | 3.381 | 0.413 | 1 | 0.521 |
| D_price | 17.470 | 2 | 0.000 | 0.251 | 0.716 | 1 | 0.397 | 5.432 | 16.753 | 1 | 0.000 |
| D_employment | 3.609 | 2 | 0.164 | -0.485 | 2.663 | 1 | 0.103 | 3.578 | 0.946 | 1 | 0.331 |
| D_wages | 3.055 | 2 | 0.217 | -0.050 | 0.028 | 1 | 0.866 | 1.966 | 3.027 | 1 | 0.082 |
| D_numFirms | 0.147 | 2 | 0.929 | -0.103 | 0.121 | 1 | 0.728 | 2.904 | 0.026 | 1 | 0.871 |
| ALL | 24.90 | 10 | 0.006 | | 3.737 | 5 | 0.588 | | 21.17 | 5 | 0.001 |

| Normality Test (Oregon) | | | | | | | | | | | |
|---|---|---|---|---|---|---|---|---|---|---|---|
| Test Type | Jarque-Bera Test | | | Skewness | | | | Kurtosis | | | |
| Equation | $\chi^2$ | Df | P-Value | Skew | $\chi^2$ | Df | P-Value | Kurtosis | $\chi^2$ | Df | P-Value |
| D_output | 4.842 | 2 | 0.089 | -0.291 | 2.796 | 1 | 0.328 | 4.171 | 3.884 | 1 | 0.049 |
| D_price | 12.403 | 2 | 0.002 | 0.376 | 1.606 | 1 | 0.205 | 4.952 | 10.797 | 1 | 0.001 |
| D_employment | 51.344 | 2 | 0.000 | -0.997 | 11.270 | 1 | 0.001 | 6.761 | 40.073 | 1 | 0.000 |
| D_wages | 12.473 | 2 | 0.002 | 0.756 | 6.447 | 1 | 0.011 | 4.455 | 5.996 | 1 | 0.014 |
| D_numFirms | 2.266 | 2 | 0.322 | -0.409 | 1.897 | 1 | 0.168 | 2.639 | 0.369 | 1 | 0.543 |
| ALL | 83.33 | 10 | 0.000 | | 22.21 | 5 | 0.001 | | 61.12 | 5 | 0.000 |

**Table B3.** Normality Tests. NAICS 321. Performed for our VEC model, these include Jarque-Bera, Skewness, and Kurtosis.

| Normality Test (Alabama) | | | | | | | | | | | |
|---|---|---|---|---|---|---|---|---|---|---|---|
| Test Type | Jarque-Bera Test | | | Skewness | | | | Kurtosis | | | |
| Equation | $\chi^2$ | Df | P-Value | Skew | $\chi^2$ | Df | P-Value | Kurtosis | $\chi^2$ | Df | P-Value |
| D_output | 39.907 | 2 | 0.000 | -1.155 | 12.445 | 1 | 0.000 | 6.431 | 27.462 | 1 | 0.000 |
| D_price | 0.093 | 2 | 0.954 | 0.089 | 0.074 | 1 | 0.785 | 3.091 | 0.019 | 1 | 0.890 |
| D_employment | 422.944 | 2 | 0.000 | -2.804 | 73.360 | 1 | 0.000 | 15.240 | 349.585 | 1 | 0.000 |
| D_wages | 32.298 | 2 | 0.000 | -0.966 | 8.708 | 1 | 0.003 | 6.180 | 23.590 | 1 | 0.000 |
| D_numFirms | 1.057 | 2 | 0.589 | 0.328 | 1.005 | 1 | 0.316 | 3.150 | 0.052 | 1 | 0.819 |
| ALL | 496.30 | 10 | 0.000 | | 95.591 | 5 | 0.000 | | 400.71 | 5 | 0.000 |

| Normality Test (Arkansas) | | | | | | | | | | | |
|---|---|---|---|---|---|---|---|---|---|---|---|
| Test Type | Jarque-Bera Test | | | Skewness | | | | Kurtosis | | | |
| Equation | $\chi^2$ | Df | P-Value | Skew | $\chi^2$ | Df | P-Value | Kurtosis | $\chi^2$ | Df | P-Value |
| D_output | 2.428 | 2 | 0.297 | -0.481 | 2.157 | 1 | 0.142 | 3.341 | 0.271 | 1 | 0.602 |
| D_price | 13.742 | 2 | 0.001 | 0.852 | 6.774 | 1 | 0.009 | 4.728 | 6.969 | 1 | 0.008 |
| D_employment | 87.085 | 2 | 0.000 | -1.665 | 25.881 | 1 | 0.000 | 8.122 | 61.205 | 1 | 0.000 |
| D_wages | 0.907 | 2 | 0.635 | -0.311 | 0.902 | 1 | 0.342 | 2.954 | 0.005 | 1 | 0.943 |
| D_numFirms | 2.310 | 2 | 0.315 | 0.497 | 2.307 | 1 | 0.129 | 3.041 | 0.004 | 1 | 0.950 |
| ALL | 106.47 | 10 | 0.000 | | 38.02 | 5 | 0.000 | | 68.45 | 5 | 0.000 |

| Normality Test (Maine) | | | | | | | | | | | |
|---|---|---|---|---|---|---|---|---|---|---|---|
| Test Type | Jarque-Bera Test | | | Skewness | | | | Kurtosis | | | |
| Equation | $\chi^2$ | Df | P-Value | Skew | $\chi^2$ | Df | P-Value | Kurtosis | $\chi^2$ | Df | P-Value |
| D_output | 9.075 | 2 | 0.011 | -0.508 | 2.796 | 1 | 0.095 | 4.523 | 6.280 | 1 | 0.012 |
| D_price | 5.445 | 2 | 0.066 | -0.374 | 1.514 | 1 | 0.218 | 4.205 | 3.931 | 1 | 0.047 |
| D_employment | 13.654 | 2 | 0.001 | -0.553 | 3.308 | 1 | 0.069 | 4.955 | 10.346 | 1 | 0.001 |
| D_wages | 1.226 | 2 | 0.542 | 0.133 | 0.191 | 1 | 0.662 | 3.618 | 1.035 | 1 | 0.309 |
| D_numFirms | 2.308 | 2 | 0.315 | -0.308 | 1.030 | 1 | 0.310 | 3.687 | 1.278 | 1 | 0.258 |
| ALL | 31.75 | 10 | 0.002 | | 8.39 | 5 | 0.183 | | 22.87 | 5 | 0.001 |

| Normality Test (Mississippi) | | | | | | | | | | | | |
|---|---|---|---|---|---|---|---|---|---|---|---|---|
| Test Type | Jarque-Bera Test | | | Skewness | | | | Kurtosis | | | | |
| Equation | $\chi^2$ | Df | P-Value | Skew | $\chi^2$ | Df | P-Value | Kurtosis | $\chi^2$ | Df | P-Value |
| D_output | 0.442 | 2 | 0.802 | -0.167 | 0.003 | 1 | 0.959 | 3.434 | 0.440 | 1 | 0.507 |
| D_price | 2.071 | 2 | 0.355 | 0.306 | 0.875 | 1 | 0.350 | 3.716 | 1.196 | 1 | 0.274 |
| D_employment | 125.104 | 2 | 0.000 | -1.671 | 26.061 | 1 | 0.000 | 9.515 | 99.043 | 1 | 0.001 |
| D_wages | 2.011 | 2 | 0.366 | 0.341 | 1.084 | 1 | 0.298 | 2.370 | 0.927 | 1 | 0.336 |
| D_numFirms | 0.431 | 2 | 0.806 | 0.184 | 0.315 | 1 | 0.575 | 2.778 | 0.115 | 1 | 0.734 |
| ALL | 130.06 | 10 | 0.000 | | 28.34 | 5 | 0.000 | | 101.72 | 5 | 0.000 |

| Normality Test (Oregon) | | | | | | | | | | | |
|---|---|---|---|---|---|---|---|---|---|---|---|
| Test Type | Jarque-Bera Test | | | Skewness | | | | Kurtosis | | | |
| Equation | $\chi^2$ | Df | P-Value | Skew | $\chi^2$ | Df | P-Value | Kurtosis | $\chi^2$ | Df | P-Value |
| D_output | 8.723 | 2 | 0.013 | -0.640 | 3.964 | 1 | 0.046 | 4.403 | 4.758 | 1 | 0.029 |
| D_price | 1.022 | 2 | 0.600 | 0.318 | 0.975 | 1 | 0.323 | 3.140 | 0.047 | 1 | 0.827 |
| D_employment | 189.101 | 2 | 0.000 | -1.985 | 38. 082 | 1 | 0.000 | 10.905 | 151.019 | 1 | 0.000 |
| D_wages | 1.780 | 2 | 0.411 | -0.273 | 0.718 | 1 | 0.397 | 3.663 | 1.061 | 1 | 0.303 |
| D_numFirms | 3.200 | 2 | 0.202 | -0.487 | 2.289 | 1 | 0.130 | 3.614 | 0.911 | 1 | 0.340 |
| ALL | 203.83 | 10 | 0.000 | | 46.028 | 5 | 0.000 | | 157.80 | 5 | 0.000 |

| Normality Test (Wisconsin) | | | | | | | | | | | |
|---|---|---|---|---|---|---|---|---|---|---|---|
| Test Type | Jarque-Bera Test | | | Skewness | | | | Kurtosis | | | |
| Equation | $\chi^2$ | Df | P-Value | Skew | $\chi^2$ | Df | P-Value | Kurtosis | $\chi^2$ | Df | P-Value |
| D_output | 9.415 | 2 | 0.009 | 0.085 | 0.067 | 1 | 0.796 | 5.002 | 9.348 | 1 | 0.002 |
| D_price | 0.720 | 2 | 0.509 | 0.216 | 0.436 | 1 | 0.509 | 3.349 | 0.284 | 1 | 0.594 |
| D_employment | 118.653 | 2 | 0.000 | -1.873 | 32.734 | 1 | 0.000 | 9.068 | 85.919 | 1 | 0.000 |
| D_wages | 2.318 | 2 | 0.128 | 0.498 | 2.317 | 1 | 0.128 | 3.012 | 0.000 | 1 | 0.985 |
| D_numFirms | 1.934 | 2 | 0.174 | 0.445 | 1.852 | 1 | 0.174 | 3.188 | 0.082 | 1 | 0.774 |
| ALL | 133.04 | 10 | 0.000 | | 37.406 | 5 | 0.000 | | 95.63 | 5 | 0.000 |

**Table B4.** Normality Tests. NAICS 322. Performed for our VEC model, these include Jarque-Bera, Skewness, and Kurtosis.

| **Normality Test (Alabama)** | | | | | | | | | | | |
|---|---|---|---|---|---|---|---|---|---|---|---|
| Test Type | Jarque-Bera Test | | | Skewness | | | | Kurtosis | | | |
| Equation | $\chi^2$ | Df | P-Value | Skew | $\chi^2$ | Df | P-Value | Kurtosis | $\chi^2$ | Df | P-Value |
| D_output | 4.283 | 2 | 0.117 | -0.529 | 2.614 | 1 | 0.106 | 3.846 | 1.669 | 1 | 0.196 |
| D_price | 0.472 | 2 | 0.790 | 0.039 | 0.014 | 1 | 0.906 | 2.557 | 0.458 | 1 | 0.498 |
| D_employment | 65.846 | 2 | 0.000 | -0.717 | 4.797 | 1 | 0.029 | 8.115 | 61.049 | 1 | 0.000 |
| D_wages | 3.412 | 2 | 0.182 | 0.455 | 1.929 | 1 | 0.165 | 3.798 | 1.484 | 1 | 0.223 |
| D_numFirms | 4.867 | 2 | 0.088 | -0.514 | 2.463 | 1 | 0.117 | 4.015 | 2.404 | 1 | 0.121 |
| ALL | 78.88 | 10 | 0.000 | | 11.816 | 5 | 0.037 | | 67.06 | 5 | 0.000 |

| **Normality Test (Arkansas)** | | | | | | | | | | | |
|---|---|---|---|---|---|---|---|---|---|---|---|
| Test Type | Jarque-Bera Test | | | Skewness | | | | Kurtosis | | | |
| Equation | $\chi^2$ | Df | P-Value | Skew | $\chi^2$ | Df | P-Value | Kurtosis | $\chi^2$ | Df | P-Value |
| D_output | 10.739 | 2 | 0.005 | 0.203 | 0.385 | 1 | 0.535 | 5.107 | 10.354 | 1 | 0.001 |
| D_price | 0.017 | 2 | 0.992 | -0.030 | 0.009 | 1 | 0.926 | 3.059 | 0.008 | 1 | 0.928 |
| D_employment | 11.551 | 2 | 0.003 | -0.842 | 6.609 | 1 | 0.010 | 4.455 | 4.942 | 1 | 0.026 |
| D_wages | 8.221 | 2 | 0.016 | 0.714 | 4.758 | 1 | 0.029 | 4.218 | 3.464 | 1 | 0.063 |
| D_numFirms | 2.270 | 2 | 0.321 | -0.464 | 2.013 | 1 | 0.156 | 3.332 | 0.257 | 1 | 0.612 |
| ALL | 32.80 | 10 | 0.000 | | 13.77 | 5 | 0.017 | | 19.03 | 5 | 0.002 |

| **Normality Test (Maine)** | | | | | | | | | | | |
|---|---|---|---|---|---|---|---|---|---|---|---|
| Test Type | Jarque-Bera Test | | | Skewness | | | | Kurtosis | | | |
| Equation | $\chi^2$ | Df | P-Value | Skew | $\chi^2$ | Df | P-Value | Kurtosis | $\chi^2$ | Df | P-Value |
| D_output | 26.231 | 2 | 0.000 | -1.265 | 14.925 | 1 | 0.000 | 5.201 | 11.306 | 1 | 0.001 |
| D_price | 0.518 | 2 | 0.772 | 0.232 | 0.502 | 1 | 0.479 | 2.917 | 0.016 | 1 | 0.898 |
| D_employment | 0.526 | 2 | 0.769 | 0.184 | 0.317 | 1 | 0.574 | 2.700 | 0.210 | 1 | 0.647 |
| D_wages | 12.719 | 2 | 0.002 | 0.331 | 1.024 | 1 | 0.312 | 5.239 | 11.695 | 1 | 0.000 |
| D_numFirms | 12.885 | 2 | 0.002 | -0.269 | 0.674 | 1 | 0.412 | 5.288 | 12.211 | 1 | 0.001 |
| ALL | 52.88 | 10 | 0.000 | | 17.44 | 5 | 0.004 | | 35.43 | 5 | 0.000 |

| Normality Test (Mississippi) | | | | | | | | | | | |
|---|---|---|---|---|---|---|---|---|---|---|---|
| Test Type | Jarque-Bera Test | | | Skewness | | | | Kurtosis | | | |
| Equation | $\chi^2$ | Df | P-Value | Skew | $\chi^2$ | Df | P-Value | Kurtosis | $\chi^2$ | Df | P-Value |
| D_output | 30.830 | 2 | 0.000 | -1.074 | 10.771 | 1 | 0.001 | 5.932 | 20.060 | 1 | 0.000 |
| D_price | 2.026 | 2 | 0.363 | -0.218 | 0.445 | 1 | 0.505 | 2.177 | 1.581 | 1 | 0.209 |
| D_employment | 5.566 | 2 | 0.062 | -0.555 | 2.874 | 1 | 0.090 | 4.074 | 2.693 | 1 | 0.101 |
| D_wages | 0.778 | 2 | 0.678 | 0.236 | 0.521 | 1 | 0.470 | 2.668 | 0.257 | 1 | 0.613 |
| D_numFirms | 1.658 | 2 | 0.436 | 0.402 | 1.506 | 1 | 0.220 | 2.744 | 0.152 | 1 | 0.696 |
| ALL | 40.86 | 10 | 0.000 | | 16.12 | 5 | 0.007 | | 24.74 | 5 | 0.000 |

| Normality Test (Wisconsin) | | | | | | | | | | | |
|---|---|---|---|---|---|---|---|---|---|---|---|
| Test Type | Jarque-Bera Test | | | Skewness | | | | Kurtosis | | | |
| Equation | $\chi^2$ | Df | P-Value | Skew | $\chi^2$ | Df | P-Value | Kurtosis | $\chi^2$ | Df | P-Value |
| D_output | 5.199 | 2 | 0.074 | -0.687 | 4.399 | 1 | 0.035 | 3.586 | 0.801 | 1 | 0.371 |
| D_price | 1.130 | 2 | 0.568 | 0.005 | 0.000 | 1 | 0.989 | 2.304 | 1.130 | 1 | 0.288 |
| D_employment | 5.491 | 2 | 0.064 | -0.553 | 2.853 | 1 | 0.091 | 4.063 | 2.638 | 1 | 0.104 |
| D_wages | 0.083 | 2 | 0.960 | 0.019 | 0.003 | 1 | 0.954 | 2.816 | 0.079 | 1 | 0.778 |
| D_numFirms | 0.083 | 2 | 0.959 | -0.094 | 0.083 | 1 | 0.774 | 2.991 | 0.000 | 1 | 0.990 |
| ALL | 11.99 | 10 | 0.286 | | 7.34 | 5 | 0.197 | | 4.65 | 5 | 0.460 |